# Complex-Dynamical Solution to the Many-Body Interaction Problem and Its Applications in Fundamental Physics

A.P. Kirilyuk

*Institute of Metal Physics, 36 Vernadsky Av., Kyiv 03142, Ukraine*

We review the recently proposed unreduced, complex-dynamical solution to the many-body problem with arbitrary interaction and its application to the unified solution of fundamental problems, including dynamic foundations of causally complete quantum mechanics, relativity, particle properties and cosmology. We first analyse the universal properties of this many-body problem solution without any perturbative reduction and show that the emerging new quality of fundamental dynamic multivaluedness (or redundance) of the resulting system configuration leads to the natural and universal concept of dynamic complexity, chaoticity and fractality of any real system behaviour. We then consider unified features of this complex dynamics and its main regimes of uniform (global) chaos and (multivalued) self-organisation, as well as the nature of physically real space, irreversibly flowing time and any system evolution in terms of its complexity conservation and transformation. Applications of that universal description to systems at various complexity levels have been performed and in this paper we review those at the lowest, fundamental complexity levels leading to causal understanding of the unified origin of quantum mechanics, relativity (special and general), elementary particles, their intrinsic properties and interactions. One reveals, in particular, the complex-dynamic origin of inertial and gravitational (relativistic) mass without introduction of any additional particle species, fields or dimensions. The related problem of the "hierarchy" of known particle masses (extreme values of Planckian units) also acquires a parsimonious solution, leading to essential modification of high-energy research strategy. Other practically important consequences and problem solutions in fundamental physics and cosmology are summarised, confirming the efficiency of that unified picture based on the complex-dynamical solution to the unreduced many-body interaction problem.

Ми даємо огляд нещодавно запропонованого нередукованого, складно-динамічного рішення задачі багатьох тіл з довільною взаємодією та його використання для одержання об'єднаного рішення проблем фундаментальної фізики, включаючи динамічне обґрунтування каузально повної квантової механіки, теорії відносності, властивостей часток та космології. Ми починаємо з аналізу універсальних властивостей рішення задачі багатьох тіл без будь-якого пертурбативного спрощення і демонструємо що виникаюча при цьому нова якість, фундаментальна динамічна багатозначність (або надмірність) виникаючої конфігурації системи, веде до природної та універсальної концепції динамічної складності, хаотичності і фрактальності поведінки будь-якої реальної системи. Далі ми досліджуємо уні-версальні властивості цієї складної динаміки та її основні режими однорідного (глобального) хаосу і (динамічно багатозначної) самоорганізації, а також динамічне походження фізично реального простору, часу що необоротно тече та еволюції будь-якої системи шляхом збереження та перетворення складності. Були реалізовані застосування такого універсального опису для систем різних рівнів складності і у цій роботі ми даємо огляд застосувань на найнижчих, фундаментальних рівнях складності які дають каузальне розуміння об'єднаної природи квантової механіки, спеціальної та загальної відносності, елементарних часток, їх властивостей та взаємодій. Зокрема, розкривається складно-динамічне походження інерційної та гравітаційної (релятивістської) маси без введення додаткових видів часток, полів та вимірів. Пов'язана проблема "ієрархії" мас відомих часток (екстремальні значення планківських одиниць) також отримує економне рішення, що веде до суттєвої модифікації стратегії досліджень у фізиці високих енергій. Ми резюмуємо також інші випливаючі практично важливі наслідки та рішення проблем у фундаментальній фізиці та космології, які підтверджують ефективність такої об'єднаної картини, заснованої на складно-динамічному рішенні нередукованої задачі багатьох тіл.

Мы даём обзор недавно предложенного нередуцированного, сложно-динамического решения задачи многих тел с произвольным взаимодействием и его использования для получения объединённого решения проблем фундаментальной физики, включая динамическое обоснование каузально полной квантовой механики, теории относительности, свойств частиц и космологии. Мы начинаем с анализа универсальных свойств решения задачи многих тел без какого-либо пертурбативного упрощения и показываем, что возникающее при этом новое качество фундаментальной динамической многозначности (или избыточности) образующейся конфигурации системы приводит к естественной и универсальной концепции динамической сложности, хаотичности и фрактальности поведения любой реальной системы. Затем мы исследуем универсальные особенности этой сложной динамики и её основные режимы однородного (глобального) хаоса и (динамически многозначной) самоорганизации, а также динамическую природу физически реального пространства, необратимо текущего времени и эволюции любой системы путём сохранения и превращения сложности. Были реализованы применения такого универсального описания для систем различных уровней сложности и в данной работе мы даём обзор применений на её нижайших, фундаментальных уровнях, дающих каузальное понимание объединённой природы квантовой механики, специальной и общей относительности, элементарных частиц, их свойств и взаимодействий. В частности, раскрывается сложно-динамическое происхождение инерционной и гравитационной (релятивистской) массы без введения дополнительных частиц, полей и измерений. Связанная проблема "иерархии" масс известных частиц (экстремальные значения планковских единиц) также получает экономное решение, что ведёт к существенной модификации стратегии исследований в физике высоких энергий. Мы резюмируем и другие вытекающие практически важные следствия и решения проблем в фундаментальной физике и космологии, которые подтверждают эффективность полученной объединённой картины, основанной на сложно-динамическом решении нередуцированной проблемы многих тел.

# 1. INTRODUCTION

The problem of many interacting bodies has no exact solution within the canonical theory framework for arbitrary interaction potential and the number of bodies greater than two. As the real world is none other than a many-body interaction process, this major feature of scholar fundamental science determines its entire "split" structure and "positivistic" result, giving rise to various rigorously incorrect, so-called "model", or "exact", solutions and their approximations within one or another version of perturbation theory. As the latter is inevitably limited to relatively small variation of a given, predetermined system configuration, usual theory cannot describe essential structure-formation processes (i.e. real change) characterised by explicit *emergence* of a *qualitatively new* object, structure and features from an essentially different, much "less structured" configuration of "interaction potential". It is such cases of explicit, essentially new (and largely varying) structure and property emergence in many-body interaction processes that represent increasingly the centre of today's practical interest, from "difficult", strong-interaction cases of the canonical many-body/solid-state problem (e.g. high-temperature superconductivity) to nanobiotechnology, genomics, brain science, pharmacology and medicine, ecology, global change problems and intelligent information and communication systems. As shown in this and other papers reviewed here, even externally different "old" problems of traditional fundamental physics (like realistic foundations of quantum mechanics, unified picture of particles and interactions, or consistent cosmology) represent but another aspect of the same underlying deficiency of absent solution to unreduced interaction problem. Long-term resistance of all these problems to applied huge efforts to rigorously, consistently solve them (already for the simplest physical systems) demonstrates the underlying basic difficulty and impossibility of its resolution within traditional, perturbative approaches and thinking.

Persisting absence of unreduced solution to already three-body interaction problem in the traditional theory framework is practically equivalent thus to the lacking understanding of real world dynamics resulting from generally strong interaction of many more than three bodies. All that remains then is a mathematically "closed" (alias "exact") and externally convenient but dramatically *incomplete*, "model" description of already appeared, "observed" structures and their small, "perturbative" variations, including "statistical physics", conventional "nonlinearity" (e.g. "solitons"), "self-organisation" ("synergetics" and "self-organised critical-ity"), "chaos" ("exponentially diverging trajectories" and "strange attractors"), "fractals" and other, *irreducibly separated* branches of scholar "complexity science" (respective references can be found e.g. in [1], see also below). The lasting stagnation and "critically" accumulating unsolved problems in solid-state physics alone, as well as its modern derivatives related to bio- and nanotechnology, underlie the urgent necessity to initiate an essential advance in consistent, causally complete (technically correct and realistic) solution of the underlying problem of *unreduced* many-body interaction going definitely beyond traditional simplification. Such solution should necessarily involve a great *qualitative novelty* with respect to all perturbative (or "exact", "statistical", "adiabatic", etc.) solutions, giving rise to the genuine, causally complete understanding, efficient design and reliable control of many-body systems at *any* level of natural and artificial (man-made) world dynamics, from elementary particles to consciousness, subject to *intense modification* by modern instrumentally powerful but intellectually deficient, always basically *empirical* technologies.

The true challenge of post-industrial science of the third millennium can be described therefore as the ultimate, reality-based extension of many-body problem and solid state physics, including now such nontraditional interacting "bodies" as universe-wide entities determining cosmological structure-formation processes, arbitrary information and communication entities (from single bits to any software and communication pieces or "agents"), various energy generating units, material and "immaterial" ("mental") brain structures, social groups and ideological "memes", ecological system and climate components, or multi-level genome and cell interactions in living organisms.

In this paper we provide, in sections 2 and 3, a creative review of the recently obtained *universal and nonperturbative* solution to the arbitrary many-body interaction problem and the related, naturally emerging and qualitatively new concept of *unreduced dynamic complexity and chaoticity* (including extended self-organisation and fractal versions) [1-6] followed, in section 4, by its applications to problems of fundamental physics and cosmology [7-19] (with only briefly mentioned links to life sciences and nanoscience [4-6,19,20]), while applications of the same complexity concept at higher levels of biology, brain science, ecology, information and communication technologies and knowledge system development need separate consideration and are presented elsewhere [4-6,21-26]. We demonstrate both the rigorous mathematical basis of this universal (and unified) science of complexity (including the clearly speci-fied qualitative novelties with respect to traditional constructions) and the resulting, essentially extended practical perspectives in fundamental physics and higher-level applications.

## 2. UNREDUCED MANY-BODY PROBLEM SOLUTION

We start with a unified, Hamiltonian formulation of arbitrary many-body interaction problem in terms of *existence equation* representing a generalised version of particular dynamic equations (e.g. Schrödinger or Hamilton-Jacobi equation) and expressing only the fact and starting configuration of unreduced (arbitrary) interaction as such:

$$\left\{ \sum_{k=0}^{N} \left[ h_k(q_k) + \sum_{l>k}^{N} V_{kl}(q_k, q_l) \right] \right\} \Psi(Q) = E\Psi(Q), \qquad (1)$$

where $h_k(q_k)$ is the generalised Hamiltonian (corresponding eventually to a dynamic complexity measure, see below, eq. (18) and section 3.2) of the $k$-th system component with the degrees of freedom $q_k$, $V_{kl}(q_k, q_l)$ is the (arbitrary) potential of interaction between the $k$-th and $l$-th components, $\Psi(Q)$ is the system state-function totally describing its configuration, $Q \equiv \{q_0, q_1, ..., q_N\}$, $E$ is the generalised Hamiltonian eigenvalue, and summations are performed over all ($N$) system components. The Hamiltonian equation form is chosen because it can be rigorously, self-consistently derived as indeed universal expression of system dynamics [1-6] corresponding to an observable measure of dynamic complexity (see below, the end of section 3.3). It is also a natural generalisation of major particular equations, such as Schrödinger equation for the wavefunction in the quantum many-body problem (solid state physics) and Hamilton-Jacobi equation for mechanical action $S$ (where $\Psi(Q) = S(Q)$) in classical mechanics. Various other equations can be reduced to a Hamiltonian form and we self-consistently confirm it and reveal the fundamental origin of that universality in our further analysis. As *physically real* time should *dynamically emerge* from our analysis (see below, section 3.2), we do not include it explicitly in the general existence equation, eq. (1), which however actually embraces the case of time-dependent potential and equation (through one of its variables, $q_k = t$) in a situation where such "formally flowing" time originates from lower complexity levels (where it has the same, universal and dynamic origin) [1-6].

We can rewrite eq. (1) in a more convenient form, reflecting the fact that one of the degrees of freedom, for example $q_0 \equiv \xi$, is physically separated from other ones, since it serves as a common, distributed system measure or interaction entity, such as position (space coordinates) or time dependence of system elements or input/output interaction field:

$$\left\{ h_0(\xi) + \sum_{k=1}^{N} \left[ h_k(q_k) + V_{0k}(\xi, q_k) + \sum_{l>k}^{N} V_{kl}(q_k, q_l) \right] \right\} \Psi(\xi, Q) = E \Psi(\xi, Q), \quad (2)$$

where now $Q \equiv \{q_1, ..., q_N\}$ and $k, l \geq 1$ here and below.

We proceed with problem expression in terms of eigenfunctions $\{\varphi_{kn_k}(q_k)\}$ and eigenvalues $\{\varepsilon_{n_k}\}$ of non-interacting components forming the necessary, known problem conditions:

$$h_k(q_k)\varphi_{kn_k}(q_k) = \varepsilon_{n_k}\varphi_{kn_k}(q_k), \quad (3)$$

$$\Psi(\xi, Q) = \sum_{n \equiv (n_1, n_2, ..., n_N)} \psi_n(\xi) \Phi_n(Q), \quad (4)$$

where $\Phi_n(Q) \equiv \varphi_{1n_1}(q_1)\varphi_{2n_2}(q_2)...\varphi_{Nn_N}(q_N)$, while $n \equiv (n_1, n_2, ..., n_N)$ runs through all eigenstate combinations (starting at $n = 0$). Inserting eq. (4) into eq. (2) and using the standard eigenfunction separation procedure (with the help of scalar product), we obtain a system of equations for $\psi_n(\xi)$ equivalent to the starting existence equation, eq. (1) or eq. (2):

$$\left[ h_0(\xi) + V_{nn}(\xi) \right] \psi_n(\xi) + \sum_{n' \neq n} V_{nn'}(\xi) \psi_{n'}(\xi) = \eta_n \psi_n(\xi), \quad (5)$$

where

$$\eta_n \equiv E - \varepsilon_n, \quad \varepsilon_n \equiv \sum_k \varepsilon_{n_k}, \quad V_{nn'}(\xi) = \sum_k \left[ V_{0k}^{nn'}(\xi) + \sum_{l>k} V_{kl}^{nn'} \right], \quad (6)$$

$$V_{0k}^{nn'}(\xi) = \int_{\Omega_Q} dQ \, \Phi_n^*(Q) V_{0k}(\xi, q_k) \Phi_{n'}(Q), \quad (7)$$

$$V_{kl}^{nn'} = \int_{\Omega_Q} dQ \, \Phi_n^*(Q) V_{kl}(q_k, q_l) \Phi_{n'}(Q). \quad (8)$$

Now one may try to solve the nonintegrable eqs. (5) by substitution of variables using the Green function technique and presumably known solutions of a reduced system of equations [1,27,28]. For that purpose, one may first separate the equation for $\psi_0(\xi)$ in the system of equations (5):

$$\left[h_0(\xi)+V_{00}(\xi)\right]\psi_0(\xi)+\sum_n V_{0n}(\xi)\psi_n(\xi)=\eta\psi_0(\xi), \qquad (5a)$$

$$\left[h_0(\xi)+V_{nn}(\xi)\right]\psi_n(\xi)+\sum_{n'\neq n} V_{nn'}(\xi)\psi_{n'}(\xi)=\eta_n\psi_n(\xi)-V_{n0}(\xi)\psi_0(\xi),$$
$$(5b)$$

where here and below $n,n'\neq 0$ and $\eta\equiv\eta_0=E-\varepsilon_0$. We try then to express $\psi_n(\xi)$ through $\psi_0(\xi)$ from eq. (5b) and substitute the result into eq. (5a). According to the well-known property of the Green function, the solution of "inhomogeneous" eq. (5b), $\psi_n(\xi)$, can be expressed through the Green function, $G_n(\xi,\xi')$, for its truncated, homogeneous part,

$$\left[h_0(\xi)+V_{nn}(\xi)\right]\psi_n(\xi)+\sum_{n'\neq n} V_{nn'}(\xi)\psi_{n'}(\xi)=\eta_n\psi_n(\xi), \qquad (9)$$

and inhomogeneous summand on the right, $-V_{n0}(\xi)\psi_0(\xi)$:

$$\psi_n(\xi)=-\int_{\Omega_\xi} d\xi' G_n(\xi,\xi')V_{n0}(\xi')\psi_0(\xi'). \qquad (10)$$

The Green function for eq. (9) is given by the standard expression:

$$G_n(\xi,\xi')=\sum_i \frac{\psi_{ni}^0(\xi)\psi_{ni}^{0*}(\xi')}{\eta_{ni}^0-\eta_n}, \qquad (11)$$

where $\{\psi_{ni}^0(\xi)\}$, $\{\eta_{ni}^0\}$ are complete sets of eigenfunctions and eigenvalues for the truncated system of equations, eq. (9). Finally, substituting eq. (11) into eq. (10) and the result for $\psi_n(\xi)$ into eq. (5a), one gets the *effective existence equation* for $\psi_0(\xi)$ alone:

$$\left[h_0(\xi)+V_{\text{eff}}(\xi;\eta)\right]\psi_0(\xi)=\eta\psi_0(\xi), \qquad (12)$$

but where the *effective (interaction) potential (EP)* $V_{\text{eff}}(\xi;\eta)$ includes the above solutions of the truncated system of equations:

$$V_{\text{eff}}(\xi;\eta)\psi_0(\xi)=V_{00}(\xi)\psi_0(\xi)+$$

$$+\sum_{n,i}\frac{V_{0n}(\xi)\psi_{ni}^0(\xi)\int_{\Omega_\xi}d\xi'\psi_{ni}^{0*}(\xi')V_{n0}(\xi')\psi_0(\xi')}{\eta-\eta_{ni}^0-\varepsilon_{n0}}, \qquad (13)$$

with $\varepsilon_{n0} \equiv \varepsilon_n - \varepsilon_0$. The eigen-solutions, $\{\psi_{0i}(\xi), \eta_i\}$, of the effective problem formulation, eqs. (12)-(13), are used then in eq. (10) to get other state-function eigen-components $\psi_{ni}(\xi)$ and obtain the total system state-function of eq. (4) (the general problem solution) in the form:

$$\Psi(\xi,Q) = \sum_i c_i \left[ \psi_{0i}(\xi) \Phi_0(Q) + \sum_{n>0} \psi_{ni}(\xi) \Phi_n(Q) \right], \quad (14)$$

where $c_i$ are coefficients to be specified by state-function matching at the boundary/configuration with zero interaction influence.

The obtained EP problem expression, eqs. (12)-(14), is but another, formally equivalent formulation of the same problem of arbitrary many-body interaction, eqs. (1)-(8). However, due to the dynamically rich environment of the unreduced EP formalism, eqs. (12)-(14), it is this problem formulation that reveals a *qualitatively new property* of a "nonintegrable" problem underlying its nonintegrability and other related properties. Namely, the strongly nonlinear EP dependence on the eigenvalues $\eta$ to be found leads to excessive, *dynamically redundant* number of problem eigen-solutions (with respect to their usually expected number), which are all *equally real* and describe *equally possible* system configurations called *realisations* that emerge as a result of one and the same interaction process development [1-7,17,28]. Being individually sufficient and therefore *mutually incompatible*, these dynamically redundant system realisations are forced, by the driving interaction itself, to *permanently replace* one another in a *dynamically random* (or *chaotic*) order thus rigorously defined. The measured system density, $\rho(\xi,Q)$, determined by squared modulus of its state-function for "wave-like" interacting entities (or state-function itself, for "particle-like" entities) is obtained then as a special, *dynamically probabilistic sum* of respective densities for all realisations:

$$\rho(\xi,Q) \equiv |\Psi(\xi,Q)|^2 = \sum_{r=1}^{N_\Re}{}^\oplus \rho_r(\xi,Q) = \sum_{r=1}^{N_\Re}{}^\oplus |\Psi_r(\xi,Q)|^2, \quad (15)$$

where $N_\Re$ is the total number of realisations, $\rho_r(\xi,Q) \equiv |\Psi_r(\xi,Q)|^2$ is the *r*-th realisation density and the *dynamically probabilistic* sum, designated by $\oplus$, describes the unceasing, *dynamically random change* of system realisations. According to eq. (14), the state-function for the *r*-th realisation, $\Psi_r(\xi,Q)$, is obtained as

$$\Psi_r(\xi,Q) = \sum_i c_i^r \Big[ \Phi_0(Q)\psi_{0i}^r(\xi) +$$

$$+ \sum_{n,i'} \frac{\Phi_n(Q)\psi_{ni'}^0(\xi) \int_{\Omega_\xi} d\xi' \psi_{ni'}^{0*}(\xi')V_{n0}(\xi')\psi_{0i}^r(\xi')}{\eta_i^r - \eta_{ni'}^0 - \varepsilon_{n0}} \Big], \quad (16)$$

where $n \neq 0$, $c_i^r$ are matching coefficients leading to causal Born's rule for realisation probabilities [1] (see below, section 3.3, eq. (37)) and $\{\psi_{0i}^r(\xi),\eta_i^r\}$ are the $r$-th realisation eigen-solutions of effective existence equation, eqs. (12)-(13).

This very important property of *dynamic multivaluedness*, or *redundance*, giving rise to *genuine but purely dynamic randomness* in *any* real many-body system (causally random realisation change) due to unreduced, *dynamically nonlinear* interaction process development (interaction feedback loops, etc.) has its inseparable partner, the property of *dynamic entanglement* of interacting system components within each system realisation. It is expressed by sums of products of eigen-functions depending on individual interacting entities $(\xi,Q)$ in eqs. (15)-(16) and is further refined by the dynamically fractal structure of the complete problem solution (see below in this section). This property (in combination with dynamic randomness) is responsible for the *tangible quality* of the resulting system "material", which is totally missing in the usual "model", perturbative theory operating only with some over-simplified, "immaterial" system image.

The number of realisations $N_\Re$ is determined by the highest power $N_{\max}$ of the characteristic equation for the efficient problem formulation, eqs. (12)-(13), straightforwardly obtained as

$$N_{\max} = N_\xi(N_q N_\xi + 1),$$

where $N_q$ and $N_\xi$ are the numbers of terms in the sums over $n$ and $i$ respectively in eq. (13). Usually $N_q = N_\xi$ and is determined by the number of initial component eigenstate combinations (see eq. (4)). Since the "ordinary" eigenvalue number of usual problem formulation of eqs. (5) (i.e. the one within each system realisation) is obviously given by $N_{q\xi} = N_q N_\xi$, it follows from the above equation that the unreduced, truly complete problem solution (eqs. (15)-(16)) contains $N_\Re = N_\xi$ such mutually incompatible, randomly changing system realisations plus one more, special system realisation consisting of an anomalously small number of only $N_\xi$ elementary eigen-solutions (instead of $N_{q\xi} = N_q N_\xi$ eigensolutions for other $N_\Re$ "regular" realisations). As confirmed by an equivalent geometric analysis [1,17] this special system realisation called *intermediate, or main, realisation* describes system configuration during its transition (and corresponding reconstruction) from one regular, "normal" realisation to another. Its anomalously small eigenvalue number reflects the fact of *transient disentanglement* of system components necessary for their new dynamic entanglement (see the previous paragraph) within the next regular realisation. Intermediate realisation provides the realistic physical basis for the *generalised wavefunction* concept [1,3,6,8,12,19,21] that corresponds, at the lowest complexity levels, to now realistically interpreted, causally complete quantum-mechanical wavefunction (see sections 3.3, 4.3).

The above transparent and purely dynamic origin of randomness within any structure (interaction result), dynamic multivaluedness, directly implies also the *dynamic origin* and *well-specified, a priori values of probabilities*, $\{\alpha_r\}$, of respective *generalised events of (observed) realisation emergence*. Indeed, as all $N_\Re$ elementary realisations have absolutely equal "rights" of appearance as a result of interaction development, one gets $\alpha_r = 1/N_\Re$, but since such elementary realisations can make dense groups of actually observed "compound" realisations, in the general case one obtains:

$$\alpha_r = \frac{N_r}{N_\Re} \left( N_r = 1,...,N_\Re; \sum_r N_r = N_\Re \right), \quad \sum_r \alpha_r = 1, \quad (17)$$

where $1 \leq N_r \leq N_\Re$ is the number of elementary realisations remaining unresolved within the $r$-th observed compound realisation. Thus dynamically determined values of *a priori* probability of realisation emergence, eqs. (17), is a natural completion of the dynamically probabilistic sum of the general solution of eq. (15). A practically important way of dynamic probability involvement is due to the generalised wavefunction of intermediate realisation mentioned above and related generalised Born's rule [1,3,6,12,19,21] connecting probability with wavefunction value (see section 3.3 for more details).

Thus rigorously derived dynamically redundant structure of manybody interaction result gives rise not only to *universally defined dynamical chaos concept* as dynamically random realisation change process, but also to closely related and equally *universal concept of dynamic complexity*, $C$, defined as any growing function of the number of system realisations or the rate of their change equal to zero for (unreal) case of only one realisation [1-6,17,19-25]:

$$C = C(N_\Re), \ dC/dN_\Re > 0, \ C(1) = 0, \qquad (18)$$

with, for example, $C(N_\Re) = C_0 \ln(N_\Re)$ or $C(N_\Re) = C_0(N_\Re - 1)$. Note that it is that unreal case of only one system realisation ($N_\Re = 1$) that is exclusively considered in usual perturbative or exact-solution approaches because of their *dynamically single-valued*, or *unitary*, structure, including scholar concepts of "chaos" and other branches of "complexity science", where fundamentally absent randomness is artificially introduced from "unknown" and then (incorrectly) "exponentially amplified" initial conditions (see [1] for more details), while complexity is defined in a non-universal, contradictory and often purely verbal way, far from the unreduced interaction problem solution, such as that of eqs. (12)-(16), clearly implied behind our *dynamic* complexity definition of eq. (18).

This complex-dynamical and chaotic structure of the unreduced many-body problem solution is further completed to its final form of *dynamically probabilistic fractal* by noting that the truncated problem solutions $\{\psi_{ni}^0(\xi)\}, \{\eta_{ni}^0\}$ entering the dynamically multivalued general solution of eqs. (12)-(16) can be further analysed by the same generalised EP method to give a hierarchy of unreduced, dynamically multivalued interaction splitting into incompatible realisations [4-6,19,21]. Indeed, applying the above Green function substitution procedure from the generalised EP method to the truncated system of equations (9), we can reformulate it as a single effective equation for $\psi_n(\xi)$:

$$\left[ h_0(\xi) + V_{\text{eff}}^n(\xi;\eta_n) \right] \psi_n(\xi) = \eta_n \psi_n(\xi) , \qquad (19)$$

where the second-level EP $V_{\text{eff}}^n(\xi;\eta_n)$ is defined as

$$V_{\text{eff}}^n(\xi;\eta_n)\psi_n(\xi) = V_{nn}(\xi)\psi_n(\xi) + \sum_{n' \neq n, i} \frac{V_{nn'}(\xi)\psi_{n'i}^{0n}(\xi) \int_{\Omega_\xi} d\xi' \psi_{n'i}^{0n*}(\xi') V_{n'n}(\xi')\psi_n(\xi')}{\eta_n - \eta_{n'i}^{0n} + \varepsilon_{n0} - \varepsilon_{n'0}} , \qquad (20)$$

and $\{\psi_{n'i}^{0n}(\xi), \eta_{n'i}^{0n}\}$ is the complete eigen-solution set for a second-level truncated system:

$$h_0(\xi)\psi_{n'}(\xi) + \sum_{n'' \neq n'} V_{n'n''}(\xi)\psi_{n''}(\xi) = \eta_{n'}\psi_{n'}(\xi), \quad n' \neq n, 0. \quad (21)$$

Similar to the first-level EP of eqs. (12)-(13), the nonlinear dependence on the eigen-solutions to be found in eqs. (19)-(20) leads to solution splitting into multiple incompatible realisations (numbered below by index $r'$), now for this first-level truncated system, eqs. (9) or (19)-(20):

$$\{\psi_{ni}^{0}(\xi), \eta_{ni}^{0}\} \to \{\psi_{ni}^{0r'}(\xi), \eta_{ni}^{0r'}\}. \quad (22)$$

Upon substitution into the first-level solution of eqs. (12)-(16), this additional, generally smaller splitting adds up with the basic system splitting into incompatible realisations, so that each first-level realisation is split into chaotically changing second-level realisations. It's easy to understand that this process continues further by splitting of the second-level truncated system, eqs. (21), by the same mechanism, leading to a yet smaller third-level truncated system, and so on, until one finally gets the simplest, integrable truncated equation for one state-function component. As a result, one obtains the *truly complete general solution* to the starting many-body interaction problem, eqs. (1), or (2), or (5), in the form of *dynamically probabilistic fractal* of observed system density $\rho(\xi,Q)$:

$$\rho(\xi,Q) = \sum_{r,r',r''...}^{N_{\Re}} {}^{\oplus} \rho_{rr'r''...}(\xi,Q), \quad (23)$$

where the dynamically probabilistic sum taken over all realisations of all levels is the final, multi-level extension of the dynamically probabilistic sum of eq. (15) accompanied by the corresponding multi-level extension of the dynamic realisation probability definition, eq. (17), now for each level of dynamically probabilistic fractal:

$$\alpha_{rr'r''...} = \frac{N_{rr'r''...}}{N_{\Re}}, \quad \sum_{r,r',r''...} \alpha_{rr'r''...} = 1, \quad (24)$$

so that the average *expectation value* of the dynamically fractal system density is given by

$$\rho_{\text{ex}}(\xi,Q) = \sum_{r,r',r''\ldots}^{N_{\Re}} \alpha_{rr'r''\ldots} \rho_{rr'r''\ldots}(\xi,Q) \ . \tag{25}$$

It is important to emphasize that the complete problem solution of eqs. (23), (24) in the form of dynamically probabilistic sum of permanently changing system realisations for the observed density distribution expresses the *exact* solution of this arbitrary, usually "nonintegrable" many-body interaction problem, rather than any reduced, perturbative expansion series. Its rigorously derived hierarchy of dynamically, *interactively* probabilistic system realisations reflects the important real-system property absent in any "model", dynamically single-valued analysis, its real-time, evolving, automatically optimal *dynamic adaptability*, where *dynamically emerging* and *probabilistically changing* system configuration in the form of the above dynamically probabilistic fractal is always optimally adjusted to external influences and all internal interactions. This important and universal property of unreduced complex dynamics can be provided with a quantitative expression showing an *exponentially huge*, practically infinite (in large systems) *power* of this *unreduced, real interaction dynamics* with respect to any its usual, dynamically single-valued model [5,6,19-25]. This maximum operation power of real, complex-dynamical system, $P$, is determined by the total number $N_{\Re}$ of its realisations that can be estimated as the *total number of combinations* of $N = N_{\text{unit}} n_{\text{link}}$ (essential) interaction links in the system (where $N_{\text{unit}}$ is the number of interacting units and $n_{\text{link}}$ the average number of links per unit):

$$P \propto N_{\Re} \simeq N! \simeq \sqrt{2\pi N}(N/e)^N \sim N^N \ . \tag{26}$$

Since in many real systems $N$ is a large number itself (for example, $N \geq 10^{12}$ for brain or genome interactions [5]), one obtains really huge, practically infinite $P$ values due to arbitrary link combinations in dynamically adaptable realisation change process. By contrast, any dynamically single-valued (basically regular and sequential) model of the same system has the power $P_0$ that can grow only as $N^\beta$ ($\beta \sim 1$), so that $P/P_0 \sim N^{N-\beta} \sim N^N \to \infty$, which clearly demonstrates the advantages of unreduced, complex-dynamic operation of a real many-body system and the related origin of "magic" properties of *living* and *intelligent* systems, while it provides also a concrete and convincing expression of strong deficiency of any its unitary (including computer) modelling [5,6,19,25].

We have thus rigorously derived a number of absolutely new properties of the unreduced, universally nonperturbative solution of arbitrary many-body interaction problem, unified under the above universally defined, reality-based and totally consistent concept of dynamic complexity, eq. (18). We also show explicitly why it cannot be obtained within any unitary, "exact" or perturbative model corresponding to zero value of this unreduced dynamic complexity (including all mechanistic "complexity" definitions within that fatally limited framework, which explains their well-known contradictions in real-world phenomena description). It is a logically transparent case of *qualitatively extended* (and now totally realistic) mathematical and physical framework: we do not neglect any part of the underlying interaction process, avoiding any its artificial "simplification", and obtain indeed a number of new, really observed qualities of the unreduced solution that could not be obtained by any means within the reduced, dynamically single-valued analysis.

The main new quality is the above dynamic redundancy, or multivaluedness, of the unreduced solution explaining many observed fundamental properties and first of all, the phenomenon of *dynamic randomness*, or *chaoticity*, that can be only incorrectly simulated in the unitary theory framework by the standard concept of "exponentially diverging trajectories" introducing no *intrinsic* randomness (which creates especially obvious problems for quantum chaos description, see section 4.3) and suffering from time-dependence of the notion of chaoticity and other key deficiencies (see [1] for more details). The related phenomenon of *fractal dynamically probabilistic entanglement* (and disentanglement) of interacting entities (degrees of freedom) within the emerging system structure determines the perceived *physical quality (texture)* of all real objects and entities, which is also absent from any usual theory "model" in the form of "immaterial" and fixed (dynamically single-valued), purely mathematical "envelopes". It is important that the universal dynamic complexity (with strictly positive and usually great value) and all its properties thus defined refer to *any real-world entity* (starting already from space and time, elementary particles and their properties, see section 4), by contrast to complexity imitations in unitary theory where ill-defined "complex systems" constitute a special class of externally "complicated" structures with many well-separated components, etc.

Many related and actively discussed but finally unclear properties of unitary description obtain now a correctly defined origin and meaning in terms of this unreduced, dynamically multivalued problem solution, eliminating any ambiguity and problems of unitary versions. These properties include nonintegrability (or "unsolvability"), noncomputability, nondecidability and various other derivatives, which all appear now as *ev-* *ident manifestations* of the *dynamically multivalued* entanglement of the unreduced problem solution (see [1,19] for discussion). It is clear also why they used to have that characteristically "mysterious", "unsolvable" air within the unitary theory framework (see e.g. [29]).

However, what is more important for us here is that these new qualities and properties of the unreduced many-body problem solution underlie the observed behaviour patterns of real systems remaining only incorrectly simulated, often unexplained and "mysterious" within usual theory models. While we shall consider some applications in detail below (section 4), it would be not out of place to emphasize the obvious and "desired" general qualities following uniquely from the above properties. They include omnipresent *dynamic uncertainty*, with its random, "undecided" switches between "competing" regimes (i.e. multiple, *incompatible* system realisations) that can be only incorrectly simulated in usual theory by standard *coexisting* "attractors" along the *same*, *single* system trajectory. This unreduced complexity manifestation is directly related to otherwise "inexplicable" behaviour of living systems, real, many-body nanosystems and various "complicated", "strong-interaction" cases of solid-state physics, including the pending high-temperature superconductivity problem. Another common feature considered in detail below (section 3.2) is due to explicitly *emerging*, or structure-forming, *events* universally explained by the underlying *realisation change process* and related real, physical *time origin* (absent in the standard, dynamically single-valued description forced to resort to artificially inserted versions of change and time). These and related features lead to the *causally complete* and *intrinsically unified* description and understanding of entire diversity of many-body interaction and structure evolution phenomena, changing completely the extension and perspectives of many-body and solid-state science, as illustrated by various applications [1,4-25] (see also section 4).

## 3. UNIVERSAL PROPERTIES OF UNREDUCED INTERACTION PROCESSES: DYNAMIC REGIMES, QUANTISATION AND CONSERVATION OF COMPLEXITY

Before proceeding to analysis of particular cases of complex dynamics of real interaction processes and resulting fundamental structures (section 4) we consider in this section *universal* properties and manifestations of unreduced interaction complexity as it is defined above (eq. (18)). It becomes evident already from the basic analysis of the previous section that natural interaction development gives rise to hierarchical, fractal structure emergence, which means that the unreduced world dynamics is organised in a hierarchy of *dynamically connected*, progressively emerging *levels of complexity*, with certain (universal) regimes of system dynamics at each level and in transition between levels.

## 3.1. Universal regimes of system behaviour: From uniform chaos to dynamically multivalued self-organisation

Each complexity level can be roughly described as unceasing system transitions between its equivalent realisations of this level taken by the system in a dynamically (and truly) random order. However, the effective "separation" (observed difference) between generic realisations with respect to characteristic values of relevant dynamical quantities (complexity measures) in each realisation can vary, determining eventually the entire diversity of observed dynamic behaviour regimes [1,2,6,17,19,21,23]. If realisations are relatively closely spaced (i.e. are similar to one another), then one obtains a relatively ordered, or *self-organised*, regime of chaos showing only small (often unobservable) random deviations from its thus well-defined average configuration. In the opposite case of relatively big difference between randomly changing realisations one will observe a strongly chaotic, explicitly irregular kind of behaviour we call here *uniform*, or *global*, *chaos*.

The origin of these qualitatively different regimes of the same, universally described interaction dynamics, as well as all the intermediate cases forming the whole variety of world dynamics, can be traced in the EP formalism expressions of the above unified formalism and in particular in the resonant structure of the key expressions for EP, eq. (13), and the system state function, eq. (16). Their resonant denominators contain the interplay between the (characteristic) separation $\Delta\varepsilon_n$ of eigenvalues $\varepsilon_{n0}$, or respective frequency $\omega_q = \Delta\varepsilon_n/\mathcal{A}_0$, of the internal dynamics of system elements (see eqs. (3), (6)), on one hand, and separation $\Delta\eta_i$ of the eigenvalues $\eta_{ni}^0$, or frequency $\omega_\xi = \Delta\eta_i/\mathcal{A}_0$, of the inter-element dynamics, on the other hand (where $\mathcal{A}_0$ is a characteristic value of generalised action, see section 3.2). If $\Delta\eta_i \ll \Delta\varepsilon_n$ (or $\omega_\xi \ll \omega_q$), then one can approximately neglect the dependence of eigenvalues $\eta_{ni}^0$ on $i$ in the denominator of EP expression, eq. (13), meaning that EP becomes local, due to completeness of the eigenfunction set $\{\psi_{ni}^0(\xi)\}$:

$$V_{\text{eff}}(\xi;\eta) = V_{00}(\xi) + \sum_n \frac{|V_{0n}(\xi)|^2}{\eta - \eta_{ni}^0 - \varepsilon_{n0}}, \qquad (27)$$

where $\eta_{ni}^0$ stands actually for the eigenvalue averaged over $i$, and we considered the driving interaction to be Hermitian. The state-function for the $r$-th realisation $\Psi_r(\xi,Q)$ from eq. (16) is simplified in a similar way:

$$\Psi_r(\xi,Q) = \sum_i c_i^r \left[ \Phi_0(Q) + \sum_n \frac{\Phi_n(Q) V_{n0}(\xi)}{\eta_i^r - \eta_{ni'}^0 - \varepsilon_{n0}} \right] \psi_{0i}^r(\xi), \qquad (28)$$

with $\eta_{ni'}^0$ being effectively averaged over $i'$.

It is this limiting case of complex dynamics that corresponds to the (generalised) self-organisation mentioned above. Indeed, it is easy to see that the effective existence equation, eq. (12), has only "ordinary" number of eigen-solutions $N_{q\xi} = N_q N_\xi$ for the local EP of eq. (27) (due to the absence of summation over $i$) corresponding to the unitary limit of only one system realisation obtained under perturbation-theory conditions (of slow inter-element and rapid intra-elements dynamics, cf. [30], section 30). It is the invariable approximation of usual self-organisation theory, or "synergetics" [31], stemming from this well-known perturbation theory case of classical dynamics. Note, however, that our generalised, *dynamically multivalued self-organisation* case has an externally, quantitatively similar but qualitatively much richer, dynamically chaotic internal structure. Indeed, even small departure from the above limiting case (finite values of $\Delta\eta_i, \omega_\xi$ for any real interaction) leads to (slightly) nonlocal EP and dynamically multivalued solution to eq. (12), implying permanent, dynamically random system realisation change ("chaotic fluctuations") around a generally well-defined ("distinct") average shape of "self-organised" system structure [1,2,6,19,21]. Whether this difference is actually observed or remains hidden under particular observation conditions, it is essential as it provides the fundamental "mode d'existence" and real origin of *any*, even quite externally "fixed" object or "regular" dynamical structure, showing that *any* real system is a *complex* one, having a well-defined complex-dynamical, *dynamically multivalued (redundant)* origin and structure, rather than only a special class of (ill-defined) "complex systems" as stated in usual, unitary-theory description.

Another important implication of the irreducible complex-dynamic structure of real, dynamically multivalued self-organisation is that it natu-rally includes another ambiguous case of usual perturbative modelling known as "self-organised criticality" (SOC) and while empirically corresponding to self-organised behaviour, remaining separated from "ordinary", "non-critical" self-organisation description and analysis. Taking into account the dynamically probabilistic and fractal, multilevel hierarchy of unreduced interaction dynamics (see the end of section 2), we can see now that *any* real distinct-shape, self-organised behaviour has an internal structure of generalised, chaotic (dynamically multivalued) SOC, in the form of permanently fluctuating "avalanches" of various sizes around the average "self-organised" system configuration. For this reason we can most correctly characterise this entire limiting case of unreduced complex dynamics as *dynamically multivalued SOC*. Note that due to the intrinsically present chaos (dynamic redundancy), it automatically resolves a usual contradiction of unitary SOC description lacking explicit chaos features (see [1,19] for further discussion and references). Similar to SOC, the unreduced, dynamically multivalued self-organisation naturally includes also other artificially "separate" cases of model (perturbative) description, such as "mode locking", "chaos control" and "synchronisation" (showing, in particular, that contrary to unitary-theory approximations, dynamic randomness can be dynamically configured but never eliminated from real system dynamics [1,2,6,19,21,23]). In a general sense, the multivalued SOC regime represents the unified complex-dynamic extension of usual dynamically single-valued *regular dynamics* constituting the essence of the entire traditional science approach.

The opposite universal case of unreduced complex (any) system dynamics is obtained when the characteristic eigenvalue separations or frequencies of system elements and inter-element dynamics are close to each other (enter in resonance), $\Delta\eta_i \simeq \Delta\varepsilon_n$ (or $\omega_\xi \simeq \omega_q$). In this case all parts of system dynamics become inseparably intermingled and cannot be separated by any approximation, while the difference between the emerging realisation configurations is relatively big (compared to characteristic parameters of each realisation). It can be seen from EP method expressions of eqs. (12)-(13) and also from their straightforward graphical analysis, which we shall not reproduce here (see refs. [1,17]). We deal therefore with the regime of *uniform, or global, chaos* characterised by maximum visible randomness of dynamic behaviour (quickly changing and essentially differing system realisations).

In order to properly characterise these two universal regimes of dynamic behaviour and transitions between them, it is convenient to intro-duce the parameter of (system) *chaoticity*, $\kappa$, determined as the ratio of the above characteristic frequencies (or eigenvalue separations) and approaching 1 in the regime of global chaos [1,2,6,19,21,23]:

$$\kappa \equiv \frac{\Delta \eta_i}{\Delta \varepsilon_n} = \frac{\omega_\xi}{\omega_q} \simeq 1 \quad . \tag{29}$$

As we have seen above, at $\kappa \ll 1$ we have the dynamically multivalued SOC (or generalised self-organisation) regime tending in the limit to quasi-total external regularity of system behaviour. With growing $\kappa$ we have progressively growing dynamic randomness of system behaviour and configuration attaining its maximum in the regime of global chaos at the main frequency resonance, $\kappa \simeq 1$. We get thus the unreduced, universally valid *meaning of the phenomenon of resonance* as the criterion of global (strongest) chaoticity of system dynamics, which extends essentially its unitary-theory understanding. The same analysis of the unreduced EP equations reveals a similar role of higher resonances as "centres of chaoticity", so that when chaoticity $\kappa$ grows from 0 (quasi-regularity) to 1 (global chaos), the degree of randomness makes a higher jump each time $\kappa$ passes through a higher resonance, $\kappa = m/n$, with integer $n > m \geq 1$. As those ever higher resonances constitute a dense network of rational values of $\kappa$, we obtain another manifestation of the "fractal structure of chaos", this time in the system parameter space.

These conclusions correlate with the well-known unitary picture of classical chaotic motion [32-35] that cannot reveal, however, the above role of resonances due to the dynamically multivalued structure of unreduced system dynamics (and its universal manifestation for *any* kind of system).[1] Note, in particular, the essential difference of the universal origin and manifestation of chaos (and order) thus revealed in our unreduced, dynamically redundant description from such major unitary-theory concepts as "overlapping resonances" (criterion of chaos), "(positive) Lyapunov exponents" (definition of chaos), or "multistability", "coexisting attractors" and "unstable periodic orbits" (structure/origin of chaos), all of these referring to a dynamically single-valued, single-trajectory behaviour (see refs. [1,19] for more details). Our unified classification of all possible regimes of any system dynamics emphasizes another essential difference from unitary complexity models: it becomes clear why and how *all* real systems/objects are complex/chaotic in their internal structure (with different proportions of randomness and order).

---

[1] We obtain, in particular, a universally applicable nonperturbative extension of the canonical perturbative KAM theory that describes the conditions of small chaoticity and absence of any essential change of system configuration. By contrast, we describe here the universal structure and origine of essential, nonperturbative changes of system configuration, i.e. unreduced (chaotic) structure formation, or emergence, phenomena.

One concrete implication of this qualitatively larger picture of our dynamically multivalued description is that we can express the above characteristic regimes of multivalued self-organisation and uniform chaos also in terms of our dynamically determined realisation probabilities $\alpha_r$, eq. (17). The uniform chaos regime with sufficiently different and quickly changing realisations corresponds to equally small probabilities of the maximum number of emerging realisations, $N_r \sim 1$ and $\alpha_r \sim 1/N_\Re$ for all $r$ in eq. (17), while an externally ordered SOC state implies a small number (usually only one) of actually observed realisations appearing with high probability but containing (contrary to unitary model description) many "invisible" realisations inside, $N_r \sim N_\Re$ and $\alpha_r \sim 1$. As system realisations are made of its original element modes "trying" to dynamically "enslave" their "competitors", this relation between "mode frequency" and "probabilistic" descriptions of possible system regimes can be approximately expressed as $\kappa \sim 1 - \alpha_r$, implying also that $\alpha_r \sim 1 - \kappa = 1 - (\omega_\xi/\omega_q)$ (for $\omega_\xi < \omega_q$).

Finally, when the chaoticity parameter passes through the global chaos value $\kappa = 1$ and then grows to infinity, we have a kind of reverse evolution of system behaviour from the highest randomness to eventual quasi-regularity (with the proper role of higher resonances), but now in an "inverse" system configuration that would normally be of less interest for a given application limited therefore to the parameter interval $0 \leq \kappa \leq 1$.

In summary, the above classification of various cases of self-organised (dynamically multivalued) and chaotic behaviour covers all possible regimes of any system existence and dynamics. We have obtained the universally valid picture of real structure emergence with changing parameters, from a highly disordered state around $\kappa = 1$ to progressively more ordered, or "self-organised" (or SOC), structure at $\kappa$ decreasing from 1 to 0, attaining finally a quasi-regular configuration for $\kappa$ around 0 (or 1). One could add here a special case of (generalised) *turbulence* emerging when one has a combination of variously ordered and chaotic regimes/structures appearing at different but *closely spaced* levels of complexity. Indeed, if the characteristic dynamical distance between neighbouring complexity levels is comparable to the dynamical distance between separate structures and regimes, then one obtains a peculiar, quickly changing coexistence of different degrees of order (SOC) and randomness (global chaos) within a single, unified dynamics. More often however the mechanism of new level formation ensures its large enough separation from neighbouring levels and thus only one dominating dynamical regime and its possible evolution with changing parameters.

**3.2. Emerging space and time hierarchy and universal conservation and transformation, or symmetry, of complexity**

Whereas the dynamically single-valued, unitary theory ($N_\Re = 1$, $C = 0$) is forced to introduce, in fact postulate, such primal notions as space and time artificially, based on their observed manifestations (including its so-called "background independent" but still postulated constructions), the dynamically multivalued description of unreduced interaction results (section 2) provides a qualitatively new possibility of *universal dynamic origin of space and time* as intrinsic features of dynamically redundant (incompatible) realisation plurality for any unreduced interaction process. Generally speaking, the inevitable *change* of incompatible but "equally real" system realisations occurring in a *dynamically random* order (section 2) gives rise to intrinsically *unstoppable* and *irreversible flow* of physically real *time* thus defined, while realisations themselves, with their physically tangible *material quality* (section 2), constitute the equally real basis for *tangible* and naturally *discrete space* structure. The naturally emerging hierarchy of interaction complexity levels gives rise to the corresponding hierarchy of physically real space and time thus defined.

Mathematically, the *space element*, or *elementary distance*, $\Delta x$, of a given complexity level is explicitly provided by the unreduced, dynamically nonlinear EP formalism, eqs. (12)-(13), as its neighbouring eigenvalue separation, $\Delta x = \Delta \eta_i^r$, where the eigenvalue separation between neighbouring realisations (numbered by $r$) gives the *elementary length* of the emerging space structure measuring typical system jump between realisations, $\lambda = \Delta x_r = \Delta_r \eta_i^r$, while the eigenvalue separation within one realisation determines the *minimum size* of real space "point" (system performing jumps), $r_0 = \Delta x_i = \Delta_i \eta_i^r$. Fundamental (dynamic) *discreteness* of thus obtained emerging space structure is due to realisation and eigenvalue discreteness of dynamically nonlinear equation of unreduced interaction formalism, while the *tangible*, "material" and physically "real" nature of this interaction-based space structure results from the property of *fractal dynamic entanglement* of interaction components within each realisation (section 2).

The *elementary time interval*, $\Delta t$, of the same complexity level is obtained as intensity or practically *frequency*, $\nu$, of causally defined *events* of incompatible *realisation emergence/change*, $\Delta t = \tau = 1/\nu$. Whereas the existence of this change and events as the necessary basis of time follows directly from the dynamic redundance of the unreduced many-body problem solution, the concrete value of $\Delta t = \tau$ can be obtained through the discrete space element $\lambda = \Delta x_r$ defined above (the length of system jump between realisations) and (known) velocity $v_0$ of signal propagation through the material of interaction components (at a lower, known level of complexity), $\tau = \lambda/v_0$. Physically real time thus defined is *unstoppably advancing* ("ticking") due to *unceasing realisation change* (driven by the interaction process itself) and it is *irreversibly* flowing due to the *causally random* choice of each next realisation (an intrinsic feature of dynamic redundance, section 2). Note especially the nontrivial link between time and causal randomness, which is inevitably ignored in the dynamically single-valued description that tends, on the contrary, to see the physical time flow as an "evident" manifestation of underlying exclusive regularity (leading to traditional problems of time irreversibility and unceasing flow). It is also important to emphasise that, contrary to space, time while being equally real is *not* a tangible material quantity but just determines the *process of change* of tangible space and therefore cannot be reasonably "mixed" with it in any reality-based "manifold" of "unified" space-time. Real unity between *emerging* space and time is of *dynamic* origin described above.

It becomes evident that multiple incompatible system realisations emerging and replacing each other as a result of its unreduced interaction process give rise to all observed (space) structures and their intrinsic change (time flow). Therefore universal dynamic complexity determined, according to eq. (18), by the total number of system realisations, $C = C(N_\Re)$, appears in various dynamical measures characterising system structure properties and evolution [1,6,13-16,19,21-23]. With elementary space and time intervals introduced above and describing system jumps between its consecutive realisations, a fundamental dynamical measure of complexity is provided by (generalised) mechanical *action*, $\mathcal{A}$, as the simplest quantity independently proportional to space and time increments:

$$\Delta \mathcal{A} = p \Delta x - E \Delta t, \tag{30}$$

where coefficients *p* and *E* are immediately recognised as (now generalised) *momentum* and (total) *energy*:

$$p = \frac{\Delta \mathcal{A}}{\Delta x}\bigg|_{t=\text{const}} \simeq \frac{\mathcal{A}_0}{\lambda} \ , \qquad (31)$$

$$E = -\frac{\Delta \mathcal{A}}{\Delta t}\bigg|_{x=\text{const}} \simeq \frac{\mathcal{A}_0}{\tau} \ , \qquad (32)$$

$\mathcal{A}_0$ being a characteristic action value at the complexity level considered (and $x$, $p$ generally understood as vectors, with partial derivatives, etc.). We see that action is an *integral* (accumulating) *measure of complexity*, while momentum and energy are related *differential* (local) *complexity measures*. In this way we obtain the universal complex-dynamical interpretation and essential extension of usual mechanical notions of action, momentum and energy to any kind of dynamics of any system. We also obtain *natural dynamic discreteness (quantisation)* of this generalised action in any real system behaviour as determined by system jumps between its discrete realisations (discrete space and time increments in eq. (30)). As these latter increments are strictly determined by the generalised EP formalism equations (section 2), i.e. by the unreduced interaction process dynamics, this quantisation of action and other quantities is very different from any formal discretisation often used e.g. in unitary computer models. As shown in section 4, the "fundamental" discreteness of quantum phenomena can also be causally explained by such quantised interaction process dynamics at the lowest complexity levels.

Because of causally irreversible time flow ($\Delta t > 0$) obtained above and positive total energy ($E > 0$, see also the next section), action-complexity is a decreasing function of time, $\Delta \mathcal{A} < 0$ (see eq. (32)). It is therefore a *consumable* form of integral dynamic complexity that is maximal at the beginning of any interaction process or system existence and then permanently goes down along its generalised "trajectory" and any transformation. We call this "potential" complexity form universally measured by the generalised action *dynamic information*, $I$ $(=\mathcal{A})$, as it represents the gradually consumed, and thus *realised*, dynamical "plan" of emerging structure formation (it should not to be confused, however, with usual information notion from computer science, etc. and can rather be considered as generalisation of usual potential energy) [1,3,16,19,21-25]. However, there is certainly another, complimentary, form of integral dynamic complexity of the same interaction process *growing* with the number of system realisations during interaction process development (see the universal definition of eq. (18)). We call the corresponding *produced* complexity measure *dynamic entropy*, $S$, as it is a generalisation of usual notion of formal statistical entropy (or, in a differential version, of usual kinetic energy). While dynamic information describes potential complexity of a structure yet to be produced, dynamic entropy is the unreduced complexity of already created structure. It follows that the decrease of the former is equal to the increase of the latter, so that their sum, the *total dynamic complexity*, $C = I + S = \mathcal{A} + S$, remains *unchanged* for any (closed) system or interaction process which is actually none other than this *unceasing transformation of dynamic information into* (the same quantity of) *dynamic entropy* preserving their sum, the total complexity $C$:

$$\Delta C = \Delta \mathcal{A} + \Delta S = 0, \ \Delta S = -\Delta \mathcal{A} > 0 . \tag{33}$$

This universal law of *conservation and transformation, or symmetry, of complexity* underlies thus *any* system, entity or process existence and dynamics, and we show (see section 3.3) that it is a universal generalisation of *all* known (correct) conservation laws and major dynamic principles [1,3,15,16,19,21-23]. In particular, since this *universal symmetry of complexity* is (exclusively) naturally *realised* in the form of *system dynamics*, there is no difference between "conservation", "transformation" and "symmetry" of complexity (contrary to unitary conservation laws, symmetries and dynamic principles). It can be considered as the unified version of "self-similarity" idea ("something cannot emerge from nothing", etc.), where we just provide the universally valid definition of this always conserved "something" for any kind of real entity or process, in the form of universal dynamic complexity (in its two forms of dynamic information and dynamic entropy).

The "unceasing transformation" part of the universal symmetry of complexity provides the ultimately general form of the "second law of thermodynamics" or "energy degradation principle", where now the permanent growth of entropy-complexity applies to *any* kind of dynamics and system, including not only arbitrary deviations from statistical equilibrium in chaotic dynamics but also inevitable internal deviations of externally regular dynamics. In other words, we have shown that due to the irreducible dynamic multivaluedness of the unreduced many-body problem solution, emergence and dynamics of any, even externally regular structure always corresponds to entropy-complexity growth (determined by emerging realisation number, eq. (18)), which resolves a persisting problem of unitary theory and reveals its fundamental origin, the dynamically single-valued reduction of multivalued real system dynamics. Correspondingly, *any* real entity resulting from interactions it contains is a *complex, dynamically multivalued* system, starting already from the simplest observable objects, elementary particles [1,7-20]. This rigorously derived conclusion is an essential extension of usual reference to a vaguely defined *special* class of "complex" (e.g. "large enough") systems contrasting with other, "non-complex" systems (that thus do not exist at all in the unreduced, dynamically multivalued picture of reality).

Another aspect of the same complexity transformation provides the universal extension of conventional "least-action principle" applicable now not only to simple "mechanical" systems but to any real system. We see now that the extended action-complexity "tends to a minimum" simply because it always decreases in favour of permanently growing entropy-complexity and that "virtual trajectories" evoked in canonical variational formulation of usual least-action principle provide a unitary imitation of quite real, plural system realisations taken by the system (and absent in dynamically single-valued models of usual theory). Therefore the "second law of thermodynamics" and the "principle of least action" are now extended to one, indivisible and absolutely universal law of conserving complexity transformation, while in usual theory they are disconnected, only empirically postulated laws applied to different system kinds.

While symmetry/conservation of system complexity is the unique, universally valid way of its existence, the corresponding *raison d'être* and *realisation* of this way is due to *inevitable internal transformation* of complexity form, from dynamic information to dynamic entropy. We obtain thus a universal complex-dynamic definition of *generalised system birth* (creation of dynamic information in the form of initial interaction configuration), *life* (spontaneous and unstoppable transformation of dynamic information into dynamic entropy, or *causally specified* unfolding of system complexity) and *death* (empty stock of dynamic information, or *generalised equilibrium*, in the form of totally unfolded complexity-entropy) [1,6,19].

### 3.3. Dynamic quantisation, "wave-particle duality" of unreduced interaction result, and universal Hamilton-Schrödinger formalism

Since we have obtained well-defined, dynamically emerging elements of physically real space and time (see the previous section), we can provide a more useful, differential-equation form of complexity conservation law by dividing eq. (33) by time increment $\Delta t|_{x=\text{const}}$:

$$\frac{\Delta \mathcal{A}}{\Delta t}\bigg|_{x=\text{const}} + H\left(x, \frac{\Delta \mathcal{A}}{\Delta x}\bigg|_{t=\text{const}}, t\right) = 0 \ , \ H = E > 0 \ , \qquad (34)$$

where the *generalised Hamiltonian*, $H = H(x,p,t)$, is a differential expression of unfolded, entropian complexity, $H = (\Delta S/\Delta t)|_{x=\text{const}}$, in agreement with the definition, eq. (32), of generalised (total) energy $E$ ($=H$) through the potential form of informational complexity-action and generalised momentum definition, eq. (31). We obtain thus the generalised, *universally* valid *Hamilton-Jacobi equation* (first part of eq. (34)) constituting a major tool of the universal formalism of unreduced dynamic complexity (see below) and revealing the true meaning of the *postulated* version of this equation from scholar classical mechanics (as well as that of usual action now generalised to dynamic complexity-action, or dynamic information, see the previous section). Note that for the case of Hamiltonian that doesn't explicitly depend on time (closed system), the Hamilton-Jacobi equation takes respective form also familiar from classical mechanics (but provided now with an extended, universal interpretation):

$$H\left(x, \frac{\Delta \mathcal{A}}{\Delta x}\Big|_{t=\text{const}}\right) = E , \qquad (34')$$

with the conserved total energy $E$ defined by eq. (32).

The condition of Hamiltonian (and total energy) positivity of eq. (34) expresses the "transformational" aspect of the universal symmetry of complexity (second part of eq. (33)) and the universal *direction/origin of the arrow of time* at all levels of complexity and corresponding time hierarchy (towards permanently growing dynamic complexity-entropy). Physically this universal time irreversibility and entropy-complexity growth is realised as *truly random* choice among *multiple* incompatible system realisations (section 3.2). This fundamental result obtains even stronger expression in terms of *generalised Lagrangian, L*, defined as the total (discrete) time derivative of informational complexity-action $\mathcal{A}$:

$$L = \frac{\Delta \mathcal{A}}{\Delta t} = \frac{\Delta \mathcal{A}}{\Delta t}\Big|_{x=\text{const}} + \frac{\Delta \mathcal{A}}{\Delta x}\Big|_{t=\text{const}} \frac{\Delta x}{\Delta t} = pv - E = pv - H , \qquad (35)$$

where $v = \Delta x/\Delta t$ is the (global) motion speed and the scalar product of vectors is implied if necessary. Intrinsic randomness of multiple realisation choice leads to the decrease of dynamic information of action (or dynamic entropy growth), eq. (33), meaning that

$$L < 0, \quad H, E > pv \geq 0 , \qquad (36)$$

which is the extended (and stronger) version of "generalised second law", or time-arrow condition, of complexity conservation of eqs. (33), (34).

The universal Hamilton-Jacobi equation for complexity-action, eq. (34), actually describes the "unfolded" system configuration made by its consecutively emerging "regular" realisations (of a certain complexity level). Details of multivalued interaction dynamics (section 2) show, however, that these are dynamically connected to each other by a special intermediate (or "main") realisation, where interacting degrees of freedom undergo transient disentanglement before entering in a new entanglement configuration within the next regular realisation. Intermediate realisation existence follows from either analytical or graphical analysis of unreduced problem solution (eqs. (12)-(13)) and its characteristic equation revealing the respective special solution with anomalously weak entanglement of interacting entities [1,3,6,12-15,19,21,22]. We call this intermediate realisation *generalised wavefunction* (or *distribution function*) $\Psi(x,t)$, as it provides the causally consistent version of the quantum-mechanical wavefunction for the corresponding (low) levels of world dynamics [1,7-20] (see also section 4.3).

An important role of the generalised wavefunction is determined by *generalised Born's rule* providing an alternative expression for realisation emergence probabilities $\alpha_r$ (see eqs. (17)) as a direct, causal consequence of interaction-driven "reduction" of intermediate realisation towards the next emerging regular realisation:

$$\alpha_r = |\Psi(X_r)|^2 \ , \qquad (37)$$

where $X_r$ is the $r$-th realisation configuration. One may also have $\alpha_r = \Psi(X_r)$ for "particle-like" complexity levels instead of eq. (37) for "wave-like" levels. Thus knowing the wavefunction one can determine realisation probabilities and the resulting system configuration without plunging into detailed calculations of elementary system realisations in the basic probability definition of eqs. (17). Hence the importance of dynamic equation for $\Psi(x,t)$ that can be derived from the same complexity conservation of eq. (34) using an additional link between the wavefunction (intermediate realisation) and complexity-action (regular realisation) based on their direct dynamical connection by generalised, causal "wave-particle duality" where the spatially extended generalised wavefunction, or "wave", of intermediate realisation evolves to the localised configuration, or "particle", of the next regular realisation.

To reveal it, note that each cycle of transition between consecutive regular realisations through the intermediate realisation can be considered as elementary act of complexity transformation between two neighbouring complexity sublevels (transiently disentangled system configuration of intermediate realisation and entangled, localised configuration of a regular realisation). As the total complexity change of a cycle should be equal to zero, while multiplicative complexity measures of sublevels are expressed by action and wavefunction respectively, one gets:

$$\Delta(\mathcal{A}\Psi) = 0, \quad \Delta\mathcal{A} = -\mathcal{A}_0 \frac{\Delta\Psi}{\Psi} \ , \tag{38}$$

where $\mathcal{A}_0$ is a characteristic action value that may also include a numerical constant depending on complexity level in question. This *dynamic quantisation* condition provides causal explanation for canonical Dirac quantisation and related wave-particle duality in quantum mechanics in terms of underlying multivalued interaction dynamics [1,8,9,11-16,19] and in general expresses the quantised structure of unreduced complex dynamics due to transitions between realisations. The quantisation condition of eq. (38) actually expresses a quasi-cyclic general character of multivalued dynamics, where the system transiently returns to the same wavefunction state after a realisation change cycle (though not without new regular realisation choice ensuring irreversible time flow and the absence of true periodicity).

Substituting now the generalised quantisation condition of eq. (38) into the universal Hamilton-Jacobi equation for complexity-action, eqs. (34)-(34'), we obtain the desired dynamic equation for the generalised wavefunction, the *generalised Schrödinger equation*:

$$\mathcal{A}_0 \frac{\Delta\Psi}{\Delta t}\Big|_{x=\text{const}} = \hat{H}\left(x, \frac{\Delta}{\Delta x}\Big|_{t=\text{const}}, t\right)\Psi(x,t) \ , \tag{39}$$

$$\hat{H}\left(x, \frac{\Delta}{\Delta x}\Big|_{t=\text{const}}\right)\Psi(x) = E\Psi(x) \ , \tag{39'}$$

where the operator form of the Hamiltonian, $\hat{H}$, is obtained from its ordinary form of eq. (34) by replacement of momentum variable $p = (\Delta\mathcal{A}/\Delta x)|_{t=\text{const}}$ with the respective "momentum operator", $\hat{p} = \mathcal{A}_0(\Delta/\Delta x)|_{t=\text{const}}$. Multivalued realisation change dynamics provides, in particular, the causal origin of quantum-mechanical Schrödinger equation (with $\mathcal{A}_0 = i\hbar$) at the corresponding lowest complexity levels [1,8,9,12-15] (see also section 4.3). Now we see, however, that this "quantum" equation has a much more general, actually universal character valid at any level of many-body world dynamics (also for "distribution functions" at "particle-like" complexity levels) and accounting for its irreducible dynamic uncertainty (multivaluedness).

Thus causally derived universal equations (34)-(39) constitute together the *unified Hamilton-Schrödinger formalism* of arbitrary (necessarily complex) many-body system dynamics [1,3,6,19,21-23] consisting of regular realisation dynamics described by the generalised Hamilton-Jacobi equation, eqs. (34), and wavefunction (or intermediate realisation) dynamics described by the generalised Schrödinger equation, eqs. (39). It is supposed that these equations should be analysed by the same unreduced EP method (section 2) which is at the origin of the underlying conservation and transformation of complexity (section 3.2). Note that by derivation these equations express the fundamental and absolutely universal symmetry of complexity (section 3.2), the single underlying law of any (complex) many-body dynamics at any level of self-developing world structure. Its "emerging" and fundamentally irreversible character is due to permanent change of multiple, mutually incompatible realisations reflected in Lagrangian negativity (or energy positivity) condition, eq.(36). This structure of the unified Hamilton-Schrödinger formalism resolves obvious contradictions of various candidate "universal" laws of complex dynamics, such as maximum entropy or maximum entropy growth rate. Whereas entropy growth remains certainly valid (in its essentially generalised form of dynamic entropy-complexity growth at the expense of dynamic information, or action-complexity) but insufficient in its canonical form for system dynamic description, maximum entropy growth rate is replaced by something like "balanced entropy growth rate", where the tendency towards quickest possible entropy growth (of new structures) is properly balanced by entropy positivity of already existing structures (interaction participants) permitting to exactly preserve the total dynamic complexity by the just right rate of its transformation from dynamic information to dynamic entropy according to universal Hamilton-Jacobi and Schrödinger equations. Those incomplete extremum principles are replaced thus by a strict-balance or universal (complexity) symmetry principle expressed by the unified Hamilton-Schrödinger formalism.

This ultimately complete and universal character of the symmetry of complexity and its expression by the Hamilton-Schrödinger equations is manifested also in the fact that the latter appear to be a unified generalisation of all known (correct) equations of linear and nonlinear science remaining separated (and usually postulated) "guesses" within ordinary theory (while the underlying universal symmetry of complexity generalises and extends all known fundamental principles from particular fields). To reveal it in an explicit form, consider a general expansion of Hamiltonian in powers of its momentum variable:

$$H(x,p,t) = \sum_{n=0}^{\infty} h_n(x,t) p^n ,  \quad (40)$$

with generally arbitrary functions $h_n(x,t)$. The unified Hamilton-Jacobi equation, eq. (34), then takes the form:

$$\frac{\Delta \mathcal{A}}{\Delta t}\Big|_{x=\text{const}} + \sum_{n=0}^{\infty} h_n(x,t) \left(\frac{\Delta \mathcal{A}}{\Delta x}\Big|_{t=\text{const}}\right)^n = 0 , \quad (41)$$

or in terms of usual "continuous" derivatives

$$\frac{\partial \mathcal{A}}{\partial t} + \sum_{n=0}^{\infty} h_n(x,t) \left(\frac{\partial \mathcal{A}}{\partial x}\right)^n = 0 . \quad (41')$$

For various series truncations and coefficients one can already reproduce here many "model" equations of usual theory, often not related to any Hamiltonian formalism (also taking into account a vector, multi-dimensional and many-body general structure of the Hamiltonian). The generalised Schrödinger equation, eq. (39), is similarly transformed into

$$\mathcal{A}_0 \frac{\Delta \Psi}{\Delta t}\Big|_{x=\text{const}} = \sum_{n=0}^{\infty} h_n(x,t) \left(\frac{\Delta}{\Delta x}\Big|_{t=\text{const}}\right)^n \Psi(x,t) , \quad (42)$$

$$\mathcal{A}_0 \frac{\partial \Psi}{\partial t} = \sum_{n=0}^{\infty} h_n(x,t) \frac{\partial^n \Psi}{\partial x^n} , \quad (42')$$

yet extending the scope of thus generalised and unified model equations. Finally, the dynamically nonlinear EP dependence on solutions to be found (see eqs. (12)-(13) in section 2) provides additional universal source of nonlinearity variously simplified in model equations and thus properly generalising their true origin.

This connection between model equations and universal formalism of unreduced dynamic complexity reveals also the origin of omnipresent "spontaneously broken symmetry" of usual theories (i.e. such a special law which is both valid and not valid), as opposed to always exact, unbroken validity of the universal symmetry of complexity (section 3.2). The latter corresponds to the unreduced, very involved in details structure of a system Hamiltonian and its further dynamic evolution according to complexity transformation. Any simplification of this structure within a model, dynamically single-valued description naturally adds the respective simplified "symmetry" but which inevitably appears to be actually "broken" upon comparison with reality of unreduced evolution of system dynamics. Only the unreduced, dynamically multivalued description provides the universal, exact and never broken symmetry of complexity leading to much more complicated dynamics and "irregular" emerging structures (as really observed in nature).

It is important to note finally that we can self-consistently confirm now universality of the Hamiltonian formalism of our starting existence equation, eq. (1), exceeding any usual "model" assumption and expressing universal symmetry (conservation) of unreduced dynamic complexity, with now properly specified origin of the Hamiltonian, energy, Lagrangian and their involved derivatives (as well as space and time variables). As demonstrated by the above expansion of eqs. (41), (42) (and dynamic nonlinearity of the generalised EP formalism, section 2), that starting Hamiltonian formalism has indeed a universal meaning exceeding its usual linear version and extending various nonlinear interaction models.

## 4. APPLICATIONS OF UNIVERSAL DESCRIPTION OF COMPLEX MANY-BODY INTERACTION DYNAMICS IN FUNDAMENTAL PHYSICS

As any structure, even the one beyond traditional "physical" reality (e.g. that of intelligence and consciousness), can only be considered as a result of interaction of its (generally simpler) constituents, the universal description of the unreduced many-body interaction process of sections 2 and 3 can be efficiently applied to such process description at different levels of world dynamics. The above analysis of universal properties and patterns of unreduced interaction shows indeed their sufficient richness and internal completeness necessary for such wide-range applications. In particular, major universal properties of unreduced interaction description are due to the key feature of fundamental dynamic multivaluedness of complete interaction results and related intrinsic chaoticity and complexity of any real structure at any level of world dynamics (contrary to artificial and inconsistent division into "complex" and "non-complex" systems in usual, dynamically single-valued description).

The resulting series of applications [1-26] starts naturally at the lowest levels of world dynamics, that of space, time, elementary particles, fields, interactions, their "intrinsic" properties and dynamic "laws" now causally, dynamically *emerging* as manifestations of the above universal features and physically real interaction complexity development [1,7-19]. It is natural that this lowest world structure level results from the simplest possible interaction configuration of two formally structureless primordial media, or "protofields", homogeneously attracted to each other (section 4.1). The next complexity level emerges as a result of interaction between these primary entities giving rise to such phenomena as causal quantum measurement, genuine quantum chaos and classical behaviour emergence in elementary closed systems that will be only briefly reviewed in this paper as transition from fundamental physics to higher-level applications.

Yet larger interaction patterns include complex nano- and biosystem dynamics (including unreduced interactive genomics) generalised to universal life properties and related medical applications [1,4-6,19,20]. Further complexity development leads to emerging (natural or artificial) intelligence and consciousness now explained as high enough levels of unreduced interaction complexity [21]. The related group of important technological applications deals with intelligent, complex-dynamic information and communication systems [23-25]. Ecological and social applications involve ever larger manifestations of unreduced dynamic complexity at the level of civilisation dynamics, including its current critical moment, now causally understood and efficiently resolved [22]. Finally, there are applications to "non-material" (but now causally understood) levels of ethical, aesthetical and spiritual entities usually studied in the humanities [1], as well as complex dynamics and qualitative transitions in (scientific) knowledge development itself [26]. Each of these higher-level applications needs a separate consideration and we limit ourselves here to a more detailed review of the unified causal solution of fundamental physics problems accumulated since the "modern physics" revolution of the twentieth century and remaining unsolved despite many efforts within the model, dynamically single-valued description.

### 4.1. Emergence of the Universe (space and time), particles and laws in a complex-dynamical interaction process

The unified hierarchy of complex-dynamical world structure (sections 2, 3) provides an extended dynamical version of the Occam's principle of parsimony, where initial structures of smaller complexity interact and give rise to ever higher-complexity structures. Correspondingly, this growing-complexity hierarchy should start from the simplest possible interaction configuration, which is obviously represented by two homogeneous material entities, called here *protofields*, homogeneously attracted to each other. This starting configuration is strongly supported also by the fact that the observed world contains two and only two long-range and omnipresent interaction forces due to gravity and electromagnetism. Therefore we identify one of the interacting media as *gravitational protofield* (with its internal degrees of freedom designated by a suitable set $\xi$) and another one as *electromagnetic (e/m) protofield* (with its internal degrees of freedom designated by a set $q$). The elementary particle structure emerging from this interaction (as specified below) shows that the physical origin of the gravitational protofield can eventually be identified as a dense (liquid-like) quark condensate, while the e/m protofield would correspond to an excited, much "lighter" (field-like) state of inter-quark (interaction) agent, such as gluon field. However, irrespective of these eventual probable interpretations, we consider those two interacting structureless protofields as a basis of emerging world structure to be rigorously derived by application of our universal interaction description.

In agreement with the above results (section 2), the starting Hamiltonian existence equation for this simplest interaction system is

$$\left[h_{\text{g}}(\xi) + V_{\text{eg}}(\xi, q) + h_{\text{e}}(q)\right]\Psi(\xi, q) = E\Psi(\xi, q) \;, \tag{43}$$

where $h_{\text{g}}(\xi)$ and $h_{\text{e}}(q)$ are the respective generalised Hamiltonians for non-interacting gravitational and e/m protofields, $V_{\text{eg}}(\xi, q)$ is their (attractive) interaction and $E$ the (generalised) energy of the resulting system configuration. It is easy to see that eq. (43) is a "condensed" version of our universal starting existence equation, eq. (2), where the "internal" interactions within protofields are included into respective Hamiltonians. Correspondingly, the whole interaction analysis of section 2 is applicable without change and we obtain the general, dynamically multivalued problem solution by the unreduced EP method in the form of eqs. (12)-(17) and related equations.

We can now analyse the structure of emerging system realisations. As we can see from eqs. (16), (13) for the state function and EP, the system in every its $r$-th (regular) realisation tends to concentrate around certain its eigenvalue $\eta_i^r$ forming thus a narrow (transient) peak of dynamically (and fractally) entangled protofields. It follows from the resonant denominator structure of both expressions, in combination with the cutting integrals in the numerator, as well as from self-amplifying dynamical link between the EP and the state function, so that the effective dynamic potential well of the former attracts additional concentration of the latter. However, due to the unavoidable realisation change process this local density peak can only be very short-living and is quickly replaced by protofield disentanglement towards the intermediate realisation of the wavefunction (quasi-free protofields) before another entanglement towards a new regular realisation concentrated around a new point of thus emerging physical space (section 3.2). This physically peaked realisation change process has a transparent physical origin in the evident instability of the homogeneously coupled protofield system with respect to local density perturbations, except the special case of pathologically high attraction force leading to a quasi-homogeneous system collapse (or rupture) [1,7,13-15,19]. As a result, for generic attraction magnitude one obtains the causally derived process of highly nonlinear local pulsation in the initially homogeneous system of attracting protofields periodically "collapsing" to *randomly* chosen points of thus *emerging* physical space. We call this process *quantum beat* and show (see below) that it totally, causally accounts for the observed quantum and relativistic behaviour, as well as intrinsic properties of thus emerging elementary *field-particles* (represented by unstoppable alternation of extended and localised states) [1,7,9-16,19]. It originates in the fundamental dynamic multivaluedness of unreduced interaction process (absent in usual dynamically single-valued "models" of many-body problem solution). Note that the total number/density of such emerging field-particles is limited by growing average tension of interacting protofields (see also section 4.4), while their main, elementary species (essentially electron and proton) are determined by respective possible deformation magnitudes ("meta-realisations") of the attracting protofields [1,15,16].

Using the general description of eqs. (31), (32) we conclude that such field-particle of the emerging first sublevel of world's complexity is characterised at rest ($p=0$) by the total energy (differential form of entropy-complexity)

$$E_0 = \frac{\mathcal{A}_0}{\tau_0} = h\nu_0 \ , \qquad (44)$$

where the characteristic value of action-complexity $\mathcal{A}_0 = h$ is naturally fixed as Planck's constant $h$ (thus obtaining a new, now causally complete interpretation [1,8-16,19]), $\tau_0$ is the period and $\nu_0$ the frequency of quan-tum beat ($v_0 \simeq 10^{20}$ Hz for the electron). At this *lowest* complexity level one starts automatically with the dynamic regime of *uniform chaos* (section 3.1, eq. (29)). The resulting *dynamically random* distribution of periodic dynamical squeeze points leads to the intrinsic property of *inertia* of quantum beat process and its energy $E_0$, appearing as system resistance to change of its *already existing* internal dynamics of *non-zero complexity* (more details below). It can also be described as chaotic wandering of the squeezed state of thus defined *virtual soliton*, in agreement with the "hidden thermodynamics" concept of Louis de Broglie [36]. We obtain also the emerging fundamental *time* measured by quantum beat period $\tau_0$ [1,8-16,19]. It is unstoppable due to the interaction-driven quantum beat pulsation between two primal entities and physically irreversible due to the truly random distribution of consecutive protofield concentration points (both features being due eventually to fundamental dynamic multivaluedness, or redundance, of unreduced interaction process).

The above state of (global) *rest* of the field-particle (i.e. its quantum beat process) of eq. (44), or actually of any isolated system, can be *rigorously defined* now as the one with minimum energy-complexity (temporal rate of system complexity transformation) and maximum homogeneity of its realisation probability distribution. Such minimum should always exist for the positively defined energy (see eqs. (34), (36)). Correspondingly, the state of (global) *motion* is defined as that with the system total energy-complexity above the minimum of the state of rest, which can only be achieved by growing inhomogeneity of realisation probability distribution (giving rise to a preferred global displacement associated with usual, empirically based motion idea) [1,8-16,19]. In the simplest case of elementary field-particle we have the totally uniform realisation probability distribution in the state of rest (uniformly chaotic wandering of the virtual soliton) and the appearing dependence of action-complexity $\mathcal{A}$ on the (emerging) space variable $x$, $\mathcal{A} = \mathcal{A}(x,t)$, in the state of motion, so that now, according to eq. (30), (35),

$$\frac{\Delta \mathcal{A}}{\Delta t} = \frac{\Delta \mathcal{A}}{\Delta t}\bigg|_{x=\text{const}} + \frac{\Delta \mathcal{A}}{\Delta x}\bigg|_{t=\text{const}} \frac{\Delta x}{\Delta t} = pv - E,$$

or

$$E = -\frac{\Delta \mathcal{A}}{\Delta t} + pv = \frac{h}{T} + \frac{h}{\lambda}v = hN + pv, \qquad (45)$$

where $E$ is the total energy of the moving system (here field-particle),

$$E = -\frac{\Delta \mathcal{A}}{\Delta t}\bigg|_{x=\text{const}} = \frac{h}{\tau} = h\nu \ , \tag{46}$$

$p$ is its universally defined momentum,

$$p = \frac{\Delta \mathcal{A}}{\Delta x}\bigg|_{t=\text{const}} = \frac{h}{\lambda} \ , \tag{47}$$

$\upsilon$ is the global motion velocity,

$$\upsilon = \frac{\Delta x}{\Delta t} = \frac{\Lambda}{T} \ , \tag{48}$$

$\tau = \Delta t|_{x=\text{const}}$ is the quantum beat (realisation change) period measured at a fixed space point, $\lambda = \Delta x|_{t=\text{const}}$ is the "quantum of space", spatial inhomogeneity emerging in the average, regular part of the moving system structure as a result of motion, $\Delta t = T$ and $\Delta x = \Lambda$ are the "total" quantum beat period and space inhomogeneity for the moving system ($\mathcal{N} = 1/T$ is the respective frequency).

It becomes clear that the motion-induced structure of a moving field-particle is none other than the famous *de Broglie wave* with the wavelength $\lambda = \lambda_\mathrm{B} = h/p$ (see eq. (47)). While the causal, dynamic wave-particle duality is a result of quantum-beat transitions between "localised" (regular) and extended (intermediate) field-particle realisations, the regular structure of de Broglie wave appears in the generally chaotic wave field of intermediate realisation (or wavefunction) due to the global motion and its inhomogeneous realisation probability distribution [1,8-16,19]. The latter global-motion tendency of de Broglie wave (second summand in the energy partition of eq. (45)) is well separated from the complementary contribution of purely random deviations from that average tendency (first summand in eq. (45)). Note, however, that every system jump between realisations (virtual soliton positions), even within the regular *average* tendency, occurs in a purely probabilistic way (due to dynamic multivaluedness), meaning that the *entire* content of total energy $E$ possesses the related property of inertia.

The unified structure of this complex field-particle dynamics implies an additional relation between its two main tendencies. If we introduce a natural definition of the speed of light $c$ as the physical velocity of perturbation propagation in the e/m protofield coupled to the gravitational protofield (causally determined by its mechanical properties), then it becomes clear that the massive particle velocity $\upsilon$ cannot exceed $c$ ($\upsilon < c$) just because of unavoidable inertia-related random deviations of virtual soliton jumps from the global motion tendency (we obtain thus a transparent causal explanation of this postulated formal limitation of standard special relativity). More exactly, we can see that during the time $\tau_1 = \lambda/c$ of global-motion advance to one de Broglie wavelength $\lambda = \lambda_B$ virtual soliton should perform $n_1 = c/v$ irregular jumps around that average motion. Since every such jump duration is $\tau$, we have $n_1\tau = \tau_1$, or $\lambda = V_{ph}\tau$, where $V_{ph} = c^2/v$ is the fictitious, formally superluminal "phase velocity" of "matter wave propagation" appearing in the original de Broglie wavelength derivation [37], where one does not take into account the dynamically random, multivalued part of internal particle dynamics. Substituting $\tau$ and $\lambda$ definitions in terms of energy and momentum, eqs. (46), (47), in the obtained relation, we get the canonical relativistic *dispersion relation* between momentum and energy:

$$p = E\frac{v}{c^2} = mv, \qquad (49)$$

where the relativistic total mass $m = E/c^2$, according to thus *rigorously substantiated* definition. In particular, for the state of rest one has $E_0 = m_0 c^2$, where $m_0$ is the dynamically defined rest mass, and the basic relation of eq. (44) can be written as

$$m_0 c^2 = h\nu_0 . \qquad (44')$$

For the general case of moving field-particle one has from eq. (46)

$$E = mc^2 = h\nu , \qquad (50)$$

which is the concise expression of extended, *causally derived* version of the famous relation between energy and mass but now revealing also the dynamic and specifically complex-dynamic (dynamically multivalued) origin of mass in the form of spatially chaotic quantum beat process (cyclic nonlinear protofield compression around a randomly chosen point alternating with extension to a quasi-homogeneous wavefunction state). Similar to energy, mass now emerges thus as differential measure of unreduced dynamic complexity (temporal rate of spatially chaotic realisation change process). Using eq. (49) in eq. (47), we finally obtain the causally derived (complex-dynamically based) canonical expression for the de Broglie wavelength:

$$\lambda = \lambda_B = \frac{h}{mv} . \qquad (51)$$

The nontrivial complex-dynamical content of externally simple eq. (49), $p = mv$, appears also through the fact that it is equivalent to now causally derived laws of "classical" Newtonian mechanics (in their relativistic version) remaining only postulated in standard theory, with only empirically defined major notions of motion, mass, energy, momentum, space and time. We can see now that even this allegedly "non-complex" dynamics laws are deeply based on the underlying dynamic complexity of unreduced many-body interaction. We can also extend these results to any complexity level for suitable cases of "smooth" enough (fine-grained) complexity structure.

Using the obtained dispersion relation of eq. (49) in the complex-dynamic energy partition of eq. (45), we arrive at the causally derived expression of time relativity revealing its true origin in the underlying complex (multivalued) interaction dynamics:

$$\tau = T\left(1 - \frac{v^2}{c^2}\right), \qquad (52)$$

where $T$ is the "internal" time period of a moving system (elementary field-particle or any other) as measured by purely random deviations of its "localised" realisations from the global motion tendency, while $\tau$ is the externally measured time period of the same moving system. We can see that the internal system time goes more slowly ($T > \tau$) because a part of a moving system complex dynamics (growing with $v$) is transferred from its internal time-making processes to those of global motion. In order to get the standard expression for relativistic time retardation with respect to the rest-frame time period $\tau_0$, we use an additional relation between $\tau$, $T$, and $\tau_0$ or the respective frequencies $\nu$, $N$, $\nu_0$:

$$N\nu = (\nu_0)^2, \quad T\tau = (\tau_0)^2. \qquad (53)$$

These relations express the physically transparent law of conservation of the total number of system realisations (as measured by reduction event frequency) due to the universal complexity conservation law (section 3.2) [1,8,13,15]. Using the second eq. (53) in eq. (52), we get the canonical expression of relativistic time retardation effect now causally explained by and derived from the underlying complex interaction dynamics:

$$T = \frac{\tau_0}{\sqrt{1 - \frac{v^2}{c^2}}}, \quad N = \nu_0 \sqrt{1 - \frac{v^2}{c^2}}. \qquad (54)$$

Other effects of special relativity, such as length contraction, are obtained as straightforward consequences of these results [1,8,13-15], with the same causal, complex-dynamic explanation behind them (see also below).

We can now provide the unified expression of complex-dynamic energy partition, eq. (45), dispersion relation, eq. (49), total energy-mass quantisation, eqs. (44), (50), and relativistic time/frequency shift, eq. (54), demonstrating the *unified causal origin of quantum and relativistic effects* in the form of underlying complex quantum beat dynamics:

$$E = h\nu_0 \sqrt{1-\frac{v^2}{c^2}} + \frac{h}{\lambda_B}v = h\nu_0\sqrt{1-\frac{v^2}{c^2}} + h\nu_B = m_0 c^2 \sqrt{1-\frac{v^2}{c^2}} + \frac{m_0 v^2}{\sqrt{1-\frac{v^2}{c^2}}},$$
(55)

where *de Broglie frequency*, $\nu_B$, is defined as

$$\nu_B = \frac{v}{\lambda_B} = \frac{pv}{h} = \frac{\nu_{B0}}{\sqrt{1-\frac{v^2}{c^2}}} = \nu\frac{v^2}{c^2},\ \nu_{B0}=\frac{m_0 v^2}{h}=\nu_0\frac{v^2}{c^2}=\frac{v}{\lambda_{B0}},\ \lambda_{B0}=\frac{h}{m_0 v}.$$
(56)

Note the ordinary relation for de Broglie wave length and frequency, $\lambda_B \nu_B = v$, confirming the physical reality of this wave but also hiding (within the above derivation) its highly nonlinear complex-dynamic (structure-formation) origin resembling a (nonlinear) "standing wave" process and naturally resolving the well-known contradictions of the canonical theory [1,8,13,15]. It is also not difficult to see [8,14,15] that $\alpha_1 = v^2/c^2$ and $\alpha_2 = 1-\alpha_1 = 1-v^2/c^2$ are the respective probabilities for the field-particle's virtual soliton to fall within the global-motion and random-deviation tendencies, in agreement with the universal realisation probability expression of eqs. (17). It confirms an even less traditional involvement of true dynamic randomness in de Broglie wave formation (global-motion tendency) and related relativistic effects.

Causally derived energy partition of eqs. (55), (56) can also be re-written as universal laws of relativistic mass and length transformation:

$$m = \frac{E}{c^2} = m_0 \left( \sqrt{1-\frac{v^2}{c^2}} + \frac{\frac{v^2}{c^2}}{\sqrt{1-\frac{v^2}{c^2}}} \right) = \frac{m_0}{\sqrt{1-\frac{v^2}{c^2}}},$$
(57)

$$\lambda_{\mathrm{B}} = \frac{\upsilon}{\nu_{\mathrm{B}}} = \frac{\upsilon\sqrt{1-\frac{\upsilon^2}{c^2}}}{\nu_{\mathrm{B}0}} = \lambda_{\mathrm{B}0}\sqrt{1-\frac{\upsilon^2}{c^2}} \ . \tag{58}$$

According to the general definition of eq. (35), the first term of complex-dynamic energy partition of eqs. (45), (55) describing the tendency of purely random virtual-soliton wandering around the global motion tendency is nothing else than the free field-particle Lagrangian $L$ (with the opposite sign):

$$L = -hN = -h\nu_0\sqrt{1-\frac{\upsilon^2}{c^2}} = -m_0c^2\sqrt{1-\frac{\upsilon^2}{c^2}} \ , \tag{59}$$

where one can see again the unified complex-dynamic origin of quantum and relativistic aspects of dynamics. We have thus consistently derived this canonical "relativistic" Lagrangian expression and provided it with a transparent physical interpretation (as accounting for purely random, "thermal" system wandering around the average global-motion tendency), contrary to formal postulation of this expression in usual theory (in addition to equally formal "principle of relativity", which is either redundant in our consistent derivation in terms of unreduced interaction dynamics) [1,8,13-15]. As noted in the general description of section 3.2, this dynamically random system wandering among its realisations provides the physically real, complex-dynamic extension of abstract "virtual trajectories" used in the canonical variational formulation of the least action principle of Lagrangian formalism.

Before proceeding to the same complex-dynamic and naturally quantised origin of "general relativity" (i.e. gravitation), let us first revise the unified dynamic nature of all "intrinsic" properties and the observed spectrum of thus emerging elementary field-particles [1,8,13-15,19]. As follows from the general analysis results of section 3.2 and those of this section for the lowest complexity level, physically real *space* emerges as a dynamically entangled and permanently chaotically "woven" combination of interacting entities, the two coupled protofields for this first complexity level. According to universal complexity conservation law (section 3.2), the number of global degrees of freedom, or "dimensions", of thus obtained tangible space should be equal to the same number for initial system configuration, i.e. two protofields plus their coupling in our case. We obtain thus the causally complete physical explanation and *origin* of the observed *number, three, of space dimensions* $N_{\mathrm{dim}}$ (or $N_{\mathrm{dim}} = n+m$ in a general case of *n* protofields coupled by *m* interactions) otherwise absent in such quality in usual unitary models. As *time* is not a material entity, it cannot constitute any similar tangible "dimension" anyhow "mixed" with spatial dimensions (unless in purely abstract models), and normally one will obtain only one global temporal variable describing *spatially chaotic* system realisation change (specified as quantum beat process). Greater than one universal temporal "lines" could exist in principle but would imply an essentially greater complexity of initial (as well as resulting) system configuration (such as a system of similar coupled systems etc.).

The observed number of (massive) elementary particle species is determined by the number of global realisations of interacting protofields depending on interaction details but basically reduced to *n* "trivial" realisations of "main" (stable) particles related respectively to (and "biased towards") each of interacting protofields. In our world ($n=2$) we obtain thus the (light) electron due to the ("elastic" and "fine") e/m protofield and (heavy) proton (eventually containing inseparable quarks) due to ("hard" and "dense") gravitational protofield (eventually a dense quark condensate). Other, less "elementary", less stable or massless particles emerge as secondary, composite and higher-level realisations.

Returning to the origin of (global) time, we note that different elementary ("main") particle species with different masses would give rise to different time rates, according to the basic law of eqs. (44'), (50), whereas in reality we observe (and widely use) the unified time flow in the whole (visible) universe (a nontrivial fact remaining unexplained and only silently postulated in usual theory). It follows from our physically real time concept that the universal time rate implies *dynamic synchronisation* of all quantum beat processes within individual "main" particles (most probably at the dominating rate of the electron). Such synchronisation is a well-known complex-dynamic phenomenon beginning here with the protofield interaction process and then "propagating" through the (coupled) e/m protofield material. It also provides a natural physical explanation for the existence of two and only two kinds of electric charge (see the next section).

The next intrinsic particle property of ambiguous physical origin within usual theory is elementary particle's spin. In our picture of quantum beat processes it naturally emerges as an inevitable nonlinear vorticity of squeezed e/m protofield (during "reduction" and then "extension" phases of quantum beat cycles) due eventually to shear instability of the locally squeezed protofield flux, by analogy to corresponding fluid motion towards a narrow outlet (but here with a much greater compressibility and dynamic nonlinearity). The same quantum beat rest energy of eq. (44) can now be also described as $E_0 = h\nu_0 = \hbar\omega_0 = h\nu_0/2 + s\omega_0$, where $\hbar = h/2\pi$, $s = \hbar/2$ is the particle spin angular momentum, while $h\nu_0/2$ and $s\omega_0$ are quantum beat energy parts due to its entangled "oscillatory" and "spinning" components respectively. We obtain a consistent explanation of the fundamental fermionic spin value (looking "anomalous" within the straightforward interpretation), its "quantised" value, direction and the origin of $\hbar$ [1,8,13-15]. Magnetic field and moment (of a particle and in general) also originate from this quantum beat vorticity (in its extended phase), in full agreement with the known electrodynamic laws [1].

**4.2. Fundamental interaction forces and related dynamic properties**

We can now proceed to the next sublevel of individual field-particle interactions naturally occurring through their common "blankets" of e/m and gravitational protofields. It becomes evident that in a system of *n* coupled protofields one will have *n* (in general *nm*) long-range interactions between individual particles through respective protofields as well as *n* accompanying short-range interactions reflecting rather lower-level interaction forces between protofield constituents (barely resolved within the emerging world dynamics). In our real case of two interacting protofields we easily identify two fundamental long-range forces with electromagnetic and gravitational ones (according to the initial system construction), while two short-range forces are identified as weak (e/m protofield) and strong (gravitational protofield) interactions thus revealing the physical nature of these usually only empirically defined forces, as well as a nontrivial connection between (the origins of) gravity and strong interaction. The number and basic properties of observed fundamental interactions are thus also causally explained within our picture, while these emerging interactions are naturally, physically *unified* by origin within unceasing quantum beat processes (see also below). The inverse square law of distance dependence of both long-range forces is evidently reproduced due to the (causally explained) three-dimensional space structure [1]. As we are dealing with interaction between dynamically discrete (periodic) quantum beat processes, both electromagnetic and gravitational interactions have the obvious *quantum origin* and are transmitted through respective protofields in the form of their deformation portions, or "quanta", that can be either better defined and quasi-stable entities (*photons* for the e/m protofield) or highly dissipative structures quickly losing their individuality (gravitational protofield excitations). Note that as all the emerging world structures are definitely biased towards its e/m protofield component (while the gravitational protofield plays the role of a heavy inertial "matrix"), we can hardly observe the detailed manifestations of all microscopic excitation processes in the gravitational protofield (contrary to macroscopic gravitation effects), in relation to respective difficulties of usual field theory, gravitational wave detection and gravity theory in general (see also below).

As the total number of fundamental interaction forces thus emerging in the system of *n* protofields with *m* coupling interactions between them is $N_F = n(m+1)$ (in the simplest case where short-range forces between protofield constituents are not related with protofield coupling forces), while the number of emerging space dimensions in the same system is $N_{\text{dim}} = n + m$ (see above, section 4.1), we obtain the following universal relations between the numbers of fundamental forces and spatial dimensions of an arbitrary causal (interaction-driven) world:

$$N_F = (m+1)(N_{\text{dim}} - m), \quad N_{\text{dim}} = \frac{N_F}{m+1} + m, \qquad (60)$$

or

$$N_F = n(N_{\text{dim}} - n + 1), \quad N_{\text{dim}} = \frac{N_F}{n} + n - 1, \qquad (61)$$

where eqs. (60) are valid for any number *n* of interacting protofields (with $m(N_{\text{dim}} - m)$ long-range and $N_{\text{dim}} - m$ short-range forces) and eqs. (61) for any number *m* of protofield coupling forces (with $n(N_{\text{dim}} - n)$ long-range and *n* short-range forces). For the simplest case where $m = 1$ the relation of eqs. (60) takes the form

$$N_F = 2(N_{\text{dim}} - 1), \quad N_{\text{dim}} = \frac{N_F}{2} + 1, \qquad (60')$$

with $N_{\text{dim}} - 1$ long-range and $N_{\text{dim}} - 1$ short-range forces. Note that these equations are interesting because they do not even depend on details of initial system configuration (*n* or *m* respectively) and can have a more general character than the underlying relations between $N_F$, $N_{\text{dim}}$ and *n*, *m*. The observed $N_{\text{dim}} = 3$, $N_F = 4$ for our world show (together with the number of "main" elementary particles $n = 2$) that $n = 2$ and $m = 1$ thus justifying the simplest starting interaction configuration between two simply coupled protofields. In general, these universal relations impose fundamental limitations on various assumptions of popular abstract models about arbitrary "mathematically convenient" numbers of "hidden dimensions", "dark matter" and other practically invisible particle species. Thus for $m = 1$ as in our world and even arbitrary (unknown) number of interacting protofields $n$, a higher dimension number, $N_{\text{dim}} > 3$, implies more diverse fundamental interaction forces, $N_{\text{F}} \geq 6$, with the evident experimentally confirmed conclusion.

The intrinsic field-particle property of *electric charge* (in the form of elementary charge $e$) is understood now as expression of long-range interaction through the e/m protofield between quantum beat processes deforming that carrier protofield. The well-known relation between $e^2$, fine-structure constant $\alpha$, the velocity of light $c$ and Planck's constant $\hbar$, $e^2 = \alpha c \hbar$, shows that electric charge (squared) expresses the same basic measure and universal quantum of *dynamic complexity* of protofield interaction as Planck's constant $h$ (a quantum of action-complexity, see above, eq. (44)). We can see now why and how electric charge is (dynamically) *quantised*. We can see also the origin of existence of exactly two "opposite" kinds of electric charge, if we take into account the property of *phase synchronisation* of all quantum beat processes ensuring the universal time flow in the universe (see above). It's evident that such synchronisation occurs up to phase inversion, so that two in-phase quantum beat processes (particles) of the *thus defined* same kind of charge will obviously repel each other (as synchronised "competitors" for the common protofield blanket), while two opposite-phase processes of *opposite* charges will naturally attract each other [1,8,13,15].

The remarkable relation between elementary charge and Planck's constant mentioned above reveals now its true meaning by providing further insight into quantum beat dynamics within elementary field-particles and the origin of fundamental constants, if we rewrite it in a slightly different form:

$$m_0 c^2 = \frac{2\pi}{\alpha} \frac{e^2}{\lambda_{\text{C}}} = N_{\Re}^e \frac{e^2}{\lambdabar_{\text{C}}}, \quad \lambda_{\text{C}} = \frac{h}{m_0 c}, \quad N_{\Re}^e = \frac{1}{\alpha}, \quad \lambdabar_{\text{C}} = \frac{\lambda_{\text{C}}}{2\pi}, \quad (62)$$

where $m_0$ is the electron rest mass and $\lambda_{\text{C}}$ the Compton wavelength. We see now that the particle rest energy $E_0 = m_0 c^2$ can be viewed as a sum of $N_{\Re}^e$ e/m interactions between two particle versions separated by a distance of $\lambdabar_{\text{C}}$. Recalling virtual soliton wandering of the underlying quantum beat dynamics (section 4.1), it means that $N_{\Re}^e$ ($= 1/\alpha = 137$) can be interpreted as the electron realisation number and $\lambdabar_{\text{C}}$ ($= \lambda_{\text{C}}/2\pi \simeq 3.9 \times 10^{-11}$ cm) as the length of elementary jump between electron realisations (both up to a numerical factor of the order of π) [1,8,13,15,16]. The latter constitutes also the basic emerging length scale at this complexity level ($\lambda = \Delta x_r$), according to the general definition of section 3.2. As to the electron reali-sation number $N_{\Re}^e$ thus derived, it provides a new, complex-dynamic interpretation of the fine-structure constant $\alpha$ remarkably coinciding (again up to a reasonable numerical factor) with the *electron realisation probability* $\alpha_r$, according to the general definition of the latter, eq. (17). Note that the Compton wavelength thus defined, eqs. (62), would correspond to the de Broglie wave length of eq. (51) for an "impossible" (massive) particle simultaneously moving with the speed of light, $v = c$, and remaining in the state of rest ($m = m_0$). However, those "impossible" properties just characterise the chaotic virtual soliton jumps (with the speed of light) *within* a massive particle at rest, thus perfectly confirming our picture and the obtained meaning of fundamental quantities.

We can further extend this new interpretation of fundamental constants by rewriting the same $e$-$\hbar$ relation in yet another form:

$$\hbar = N_{\Re}^e \frac{e^2}{c} = \lambdabar_C p_0, \quad \lambdabar_C = N_{\Re}^e r_e, \tag{63}$$

where $p_0 = m_0 c = E_0/c$ and $r_e = e^2/m_0 c^2$ ($\simeq 2.8 \times 10^{-13}$ cm) is the usual "classical radius" of the electron. As we deal with the EP (section 2) of interacting protofields realised in each quantum beat process, we can see from the first of eqs. (63) that $N_{\Re}^e$ or $\lambdabar_C$ can be interpreted as this EP width, while $e^2/c$ or $p_0$ its respective depth and $\hbar$ its universal "volume" (the same for all particle species and their coherent agglomerations). Universality of EP volume and thus of Planck's constant (remaining a postulated "quantum mystery" within all usual approaches) follows thus from the general complexity conservation law ($h$ being the lowest-level quantum of action-complexity) and results physically from permanent protofield properties and coupling strength for all eventually emerging realisations. We obtain also another meaning of the fine-structure constant $\alpha$ as a quantity inversely proportional to the EP width for the electron ($\alpha = 1/N_{\Re}^e$) and thus proportional to its depth or e/m interaction strength. The second of eqs. (63) shows also that the EP width $\lambdabar_C$ contains exactly $N_{\Re}^e$ sizes of $r_e$, which together with realisation set completeness implies that $r_e$ determines the size of each regular (localised) realisation of virtual soliton or emerging physical space "point" ($r_0 = \Delta x_i$), in agreement with the general definition of section 3.2. Another estimate of the virtual soliton size implies that localised realisations should densely fill in a circle with the radius $\lambdabar_C$ of a single jump length (around a jump starting point), giving the virtual soliton size of $2\pi r_e$ ($= 2\pi \lambdabar_C / N_{\Re}^e$).

In the whole we obtain therefore a universal, consistent and causal interpretation of the origin, role and conservation of major physical con-stants and intrinsic particle properties now unified also with equally causal description of quantum and relativistic behaviour liberated from usual postulated "mysteries" and formal definitions. Various massive particle species emerge as different realisations of local quantum beat dynamics, varying from the relatively large and shallow EP for the electron ($N_\Re = N_\Re^e \simeq 137$, $\alpha = 1/N_\Re^e \ll 1$) to the deep and narrow EP for the heaviest observed particles ($N_\Re, \alpha \sim 1$), while the volume of all respective EP wells remains the same and equal to $\hbar$. As the same protofield dynamics remains valid also for dense, wave-coherent ("quantum") particle agglomerates, we obtain a physically transparent explanation of approximate equality between the largest individual-particle mass and the heaviest nuclear mass (around few hundred GeV, up to variations for very unstable species) [9,11,15,16]. The nuclei roughly behave thus as large "elementary" particles with a particularly complex internal dynamics.

On the other hand, massless photons are explained as small enough deformations of the coupled e/m protofield, which are not sufficient to overcome its average tension and produce its localised "reduction" giving rise to spatially chaotic quantum beat and finite inertia. However, they do weakly interact with the underlying gravitational protofield losing energy to its internal degrees of freedom, which provides a causal explanation for the "red shift" effect over long propagation distances without usual Big Bang expansion [15,16] (see also section 4.4 for more details on complex-dynamical cosmology). "Virtual" interaction-exchange and "zero-point" photons only formally introduced in usual theory become now quite real (and naturally quantised) e/m protofield perturbations of the same origin, eventually due to quantum beat dynamics of massive (interacting) particles. By contrast, the existence of *massive* "zero-point" (but not exchange) virtual particles would be very improbable in our description being forbidden by the fundamental complexity conservation law (because mass is the unreduced dynamic complexity measure), which has important consequences for the "cosmological constant problem" (actually solved now) and related cosmology picture (section 4.4).

The second long-range interaction force between particles emerges as their quantum beat interaction through the gravitational protofield and is observed as universal gravitation. It would be similar to analogous e/m interaction through the e/m protofield except that the world's reality is strongly displaced towards the actually structure-forming and much more deformable e/m protofield, while the dense and dissipative gravitational protofield remains a usually directly unseen (though quite real) material background. A major related feature is the existence of only one, attractive kind of gravitational interaction as temporal phases of interacting quantum beat processes and respective exchange perturbations (giving rise to world-wide synchronisation and two kinds of charge and force in the case of e/m protofield) are not either observable or even preserved within a highly dissipative (quark-condensate) matter of the gravitational protofield. Any two quantum beat processes are simply attracted to each other through respective density changes of the common gravitational protofield blanket, and although thus explained gravitational attraction has a *naturally quantised origin* (cf. respective problems in usual theory), it can hardly be presented as occurring through "exchange of gravitons" (by analogy to exchange of photons through the e/m protofield), simply because contrary to photons gravitational protofield perturbations cannot preserve their individuality over any reasonably large distances in the highly dissipative environment of the gravitational protofield condensate. Gravitational field itself around a massive particle is physically realised as a change of the gravitational protofield tension/density due to particle's quantum beat "squeezing" action.

As the average gravitational protofield deformation giving rise to gravitational interaction grows with inertial mass defined above as quantum beat temporal rate, eqs. (44), (50), we obtain the *physically substantiated* (rather than formally postulated) principle of equivalence between gravitational and inertial mass and the Newtonian gravity law (gravity force proportional to inertial masses) for weak fields. Note, however, that gravitational and inertial manifestations of (relativistic) mass-energy are its related but qualitatively different aspects, very far from formal coefficient identity in a unitary theory.

Within the *same*, dynamically multivalued system dynamics we also obtain equally *physically* emergent effects of "general relativity" in addition naturally unified with the above quantum origin of gravity. Thus the key effect of time retardation in gravitational field results from a physically evident change of the gravitational protofield tension/density (eventually resulting from massive particles creating that gravitational field). As this (modified) gravitational protofield density determines the local quantum beat frequency for a test particle (determining the physically real time flow), we obtain instead of eq. (50):

$$M(x)c^2 = h\nu(x) = mc^2\sqrt{g_{00}(x)}, \qquad (64)$$

where $\nu(x)$ is the local quantum beat frequency of a (generally moving)

test particle, $m$ is its relativistic mass in the absence of gravitational field, and the "metric" $g_{00}(x) < 1$ describes in reality the locally modified gravitational protofield density. In weak fields $g_{00}(x) = 1 + 2\phi_g(x)/c^2$, where $\phi_g(x) < 0$ is the gravitational field potential [38]. As $v(x)$ determines the local time flow rate, we obtain the physical origin of experimentally confirmed time retardation in gravitational field [1,8,9,13-15], without any formal postulates and reference to geometric deformation of a formal mixture of abstract time and space variables (though it could be used, in principle, as a formal description framework within respective limitations, including the obvious difficulty of such "geometric" gravity quantisation). Other effects of general relativity (e.g. light ray "bending" in gravitational field) can be reproduced in the same physically transparent way as being due to gravitational protofield inhomogeneities induced eventually by massive quantum beat processes.

　　　Having obtained thus the emerging, physically real and unified complex-dynamic nature of intrinsic particle properties, their "relativistic" and "quantum" behaviour and all the four fundamental interaction forces, we can return to the detailed physical unification of these interaction forces and related Planckian unit, or mass hierarchy, problem. Usual values of Planckian units are obtained by purely dimensional, formal combination of Planck's constant $\hbar$ (for quantum aspects), the speed of light $c$ (for e/m and special relativistic aspects) and the gravitational constant $\gamma$ from classical Newton's law (for gravitational and general relativistic aspects). As a result one obtains the characteristic Planckian units of length $l_P$, time $t_P$ and mass $m_P$, which have however too extreme values separated by many orders of magnitude from any observable or even conceivable (necessary) values for any extreme particle species (thus $m_P$ attains almost macroscopic mass values). This "mass hierarchy" problem remains basically unsolved in usual theory (without introduction of purely abstract and contradictory "invisible dimensions" in brane-world modifications), which persists in using these conventional extreme values of Planckian units for its major models and essential results ("standard model", cosmology scenarios, quantum gravity, etc.).

　　　In our physically connected description the origin of this problem becomes immediately evident: while two of the used constants, $\hbar$ and $c$, are directly related to the interacting protofield properties and quantum beat dynamics, the third constant, $\gamma$, describes only an indirect, long-range interaction between two particles though the deeper, directly imperceptible gravitational protofield. In fact, the "genuine" Planckian units represent now not simply formal dimensional combinations but the observed parameters of the real quantum beat process for an extreme (the heaviest) field-particle possible for a given protofield interaction. Therefore the usual gravitational constant $\gamma$ coming from the indirect long-range interaction, should be replaced for these genuine units by an effective short-range constant $\gamma_0 \gg \gamma$ directly characterising the intrinsic gravitational protofield properties and therefore strongly exceeding the conventional indirect-interaction value weakened by a "long" interaction-transfer process between protofields. This short-range $\gamma_0$ value can be interpreted as characterising zero-distance gravitational interaction practically "within" the field-particle, where it is effectively, dynamically unified with all other interactions being reduced to the maximum local magnitude of protofield interaction within the virtual-soliton (maximum-squeeze) state. Substituting $\gamma_0$ for $\gamma$ in usual expressions, we obtain thus the new, "renormalized" values of Planckian units ($L_P$, $T_P$, $M_P$) now corresponding to *observed* (extreme) properties ($l_{exp}$, $t_{exp}$, $m_{exp}$) of the heaviest particle and its (most intense) quantum beat process:

$$L_P = \sqrt{\frac{\gamma_0 \hbar}{c^3}} \simeq 10^{-17} - 10^{-16} \text{ cm} = l_{exp} \; ,$$

$$T_P = \sqrt{\frac{\gamma_0 \hbar}{c^5}} \simeq 10^{-27} - 10^{-26} \text{ s} = t_{exp} \; , \quad (65)$$

$$M_P = \sqrt{\frac{\hbar c}{\gamma_0}} \simeq 10^{-22} - 10^{-21} \text{ g } (10^2 - 10^3 \text{ GeV}) = m_{exp} \; ,$$

where the relation between $\gamma_0$ and $\gamma$ can be determined, for example, from the one between usual Planckian unit of mass and the observed largest particle mass (up to its evolving value): $\gamma_0 = (m_P/m_{exp})^2 \gamma \simeq (10^{33} - 10^{34})\gamma$. The hierarchy problem of the observed mass spectrum is thus naturally resolved in the physically transparent and parsimonious way (without new, "hidden" entities introduction) further completing the entire, already intrinsically unified picture. We solve simultaneously the related problem of particular weakness of gravitational interaction as being due to the small value of usual, long-range value of gravitational constant related to (weak) interaction transfer between protofields.

**4.3. Interacting particle dynamics and classical behaviour emergence**

According to the general process of complexity unfolding from dynamic information to dynamic entropy (section 3.2) in successive emergence of each next complexity (sub)level from unreduced interaction of entities of lower levels, we can now proceed to always rigorously derived description of the next sublevel of interacting (quantum) field-particles. In this case we need only to specify the universal Hamilton-Schrödinger formalism of section 3.3. While the "average" classical trajectories are described by the universal Hamilton-Jacobi equation, eqs. (34), (34'), or more directly by extended Newton's second law resulting from relativistic (now complex-dynamically derived) dispersion relation of eq. (49), the wave dynamics of intermediate realisation of the wavefunction is more important at this essentially quantum complexity level of relatively big chaotic particle jumps between realisations. It starts from the quantisation relation of eq. (38) now specified as

$$\Delta \mathcal{A} = -i\hbar \frac{\Delta \Psi}{\Psi} \ , \tag{66}$$

where $\mathcal{A}_0 = i\hbar$ is the characteristic action coefficient defined by the absolutely universal value of Planck's constant at this lowest, "indivisible" complexity level. Combining it with the universal momentum and energy definitions of eqs. (31), (32) (or eqs. (46), (47)) and using the continuous versions of partial derivatives (at these small scales), we obtain the canonical Dirac quantisation rules:

$$p = \frac{\Delta \mathcal{A}}{\Delta x}\bigg|_{t=\text{const}} = -\frac{1}{\Psi} i\hbar \frac{\partial \Psi}{\partial x} \ , \quad p^2 = -\frac{1}{\Psi} \hbar^2 \frac{\partial^2 \Psi}{\partial x^2} \ , \tag{67}$$

$$E = -\frac{\Delta \mathcal{A}}{\Delta t}\bigg|_{x=\text{const}} = \frac{1}{\Psi} i\hbar \frac{\partial \Psi}{\partial t} \ , \quad E^2 = -\frac{1}{\Psi} \hbar^2 \frac{\partial^2 \Psi}{\partial t^2} \ , \tag{68}$$

where the higher powers of $p$ and $E$ properly reflect the wave nature of $\Psi$ [1,12] and vectors can be naturally assumed where necessary. Note that these quantisation rules, only formally postulated in usual quantum theory, are now causally derived as the direct description of realisation change within the physically real cycle of quantum beat process (between the extended wavefunction and localised virtual soliton states) always preserving globally the same system state. The same refers to the related formalism of "production and annihilation operators" describing the alternating "pro-duction" and "annihilation" events of now physically real "corpuscular", localised states of participating field-particles (or larger "coherent" entities in general) [1].

In agreement with the general theory of section 3.3, application of quantisation rules to the Hamilton-Jacobi equation (for localised states) gives the Schrödinger equation for the wavefunction (cf. eq. (39)):

$$i\hbar \frac{\partial \Psi}{\partial t} = \hat{H}\left(x, -i\hbar \frac{\partial}{\partial x}, t\right)\Psi(x,t) \ , \qquad (69)$$

or, for the simplest interaction Hamiltonian, $H(x,p,t) = p^2/2m + V(x,t)$,

$$i\hbar \frac{\partial \Psi}{\partial t} = -\frac{\hbar^2}{2m} \frac{\partial^2 \Psi}{\partial x^2} + V(x,t)\Psi(x,t) \ . \qquad (70)$$

It is important to emphasize that the Schrödinger equation thus rigorously derived from the underlying complex (multivalued) interaction dynamics (of the system of two protofields) describes the evolution of a *physically real* wavefunction permanently alternating, however, with the chaotically selected localised states of the virtual soliton (where this chaoticity is at the origin of mass entering the equation). It is accompanied by equally causally derived (and now universal for all complexity levels) Born's rule of eq. (37) reflecting the physically real transformation of extended wavefunction to the "reduced" state of virtual soliton during each quantum beat cycle (eventually included into a higher-level measurement process):

$$\alpha(x,t) = |\Psi(x,t)|^2 \ , \qquad (71)$$

where $\alpha(x,t)$ is the probability of finding the particle at the point *x* at the time moment *t*. Therefore we don't need to artificially introduce any additional, externally originating "decoherence" or "collapse" processes in the Schrödinger equation (remaining always exact) or in related measurement processes (see also below), in contrast to various attempts of such mechanistic insertion of a necessary (but actually never causal) source of randomness and localisation in "decoherence" and "dynamical collapse" interpretations of observed quantum behaviour (e.g. [39-45]).

This genuine, complex-dynamic origin of the Schrödinger equation provides also a much deeper physical meaning of its bound state discreteness (including the non-zero ground-state energy) as finally originating not in the formal mathematical "standing-wave discreteness" in a binding potential well but in the underlying quantum beat dynamics, so that those observed "standing waves" are in reality produced and maintained by permanent events of (highly nonlinear) wavefunction reduction to compact virtual soliton states at the global wave nodes [1]. That observed global, "nonrelativistic" tendency is accompanied by a great deal of purely random, "relativistic" virtual soliton wandering around it accounting for particle's mass as well as for the "quantum tunnelling" effect thus causally explained now [1,17].

In order to obtain fully relativistic wave equations, one can insert the causally derived quantisation relations, eqs. (66), (67), into the causal relativistic energy partition of eq. (55) (of the same complex-dynamic origin) rewritten as

$$E = m_0 c^2 \sqrt{1 - \frac{v^2}{c^2}} + \frac{p^2}{m} \ , \quad \text{or} \quad E^2 = m_0^2 c^4 + p^2 c^2 \ ,$$

which gives the Klein-Gordon or Dirac equation for a free particle:

$$\frac{\partial^2 \Psi}{\partial t^2} - c^2 \frac{\partial^2 \Psi}{\partial x^2} + \omega_0^2 \Psi = 0 \ , \tag{72}$$

where $\omega_0 = m_0 c^2 / \hbar = 2\pi \nu_0$ is the "circular" frequency of the rest-frame quantum beat actually accounting for its causally explained spin vorticity (see above, section 4.1). More elaborated forms of relativistic wave equation taking into account particle interactions can be derived by a similar causal quantisation procedure [1]. In the nonrelativistic limit they are reduced to the Schrödinger equation, eq. (70), already obtained above.

One can also mention here the causal dynamic origin of a specific quantum "interaction" effect known as *quantum entanglement* in many-particle systems and constituting a classic "quantum mystery" as if hinting on possible "nonlocal interaction" between separated particles occurring at arbitrary high speed of interaction transmission. In the absence of properly specified complex-dynamic origin of physical particle-processes in usual theory, the experimentally observed "quantum correlations" of separated quantum particles entering the system of major quantum postulates will indeed appear as "inexplicable". On the other hand, our unified picture of underlying quantum beat dynamics within each elementary particle accounting for all quantum and relativistic properties (as described above in section 4) provides also a natural explanation of those quantum correlations at a distance as *phase synchronisation* between "coherent" (i.e. "quantum") system components [12,19]. As noted in section 4.2, such *temporal* phase synchronisation (up to phase inversion) has a global character accounting for existence of two "opposite" kinds of electric charge throughout the universe as well as the universal flow of its physically real time. However, quantum correlations in closer or specially prepared many-particle systems would often need more detailed temporal and spatial coherence of individual quantum beat (and photon-oscillation) processes that readily occurs due to properly organised interaction and provides a direct and simple explanation of quantum correlations at a distance, without any additional assumption or supernatural mystification. In a similar way, synchronised quantum jumps of interacting quantum beat processes provide a causal dynamic origin for the Pauli exclusion principle and other canonically postulated rules for many-body fermionic and bosonic particle systems, in relation to the causal origin of these two kinds of particle behaviour themselves (see [1,12,19] for more details which we won't consider here).

We can only briefly mention major phenomena of further dynamic complexity development for systems of interacting quantum particles whose detailed consideration needs a separate review. They include genuine quantum chaos [1,15,17,19], causal quantum measurement dynamics [1,18] and classicality emergence in elementary bound (isolated) particle systems [1,8,9,13-15,19].

In the case of *quantum chaos* [1,15,17,19] we deal with a (generally many-body) quantum interaction problem with the Hamiltonian (nondissipative) dynamics described e.g. by the Schrödinger equation such as eq. (70) with an arbitrary, "nonintegrable" interaction configuration (i.e. practically more complicated than a particle in a one-dimensional time-independent potential). The well-known persisting difficulty of usual (dynamically single-valued) quantum chaos theory is that it cannot simulate genuine dynamic randomness by "exponentially diverging trajectories" of its classical counterpart because of the absence of any well-defined quantum trajectories and "smearing" effect of (regular) quantum discreteness (while the notorious "quantum uncertainty" remains a separate, formally postulated and measurement-related feature). Therefore even classically chaotic interaction configurations seem to produce only regular quantum dynamics, in strong contradiction to the fundamental correspondence principle of transition between quantum and classical dynamics in the limit of $\hbar \to 0$. Our analysis with the help of unreduced, dynamically multivalued solution of a standard quantum chaos problem shows [1,17,19] that this fundamental difficulty does not appear within this complete problem solution providing its universal origin of purely dynamic randomness (section 2), in full agreement with the canonical correspondence principle now extended to chaotic systems, while the origin of usual theory difficulty is re-vealed as its hugely restricted dynamically single-valued model. The quantum chaos case provides therefore a particularly transparent demonstration of qualitative advantages of extended, dynamically multivalued interaction description (that may remain more hidden and subject to misleading imitations in the "fine-grained" structure of classical chaos [1] or conventional "postulated" randomness of quantum measurement). Both the general quantum chaos analysis and its global chaos criterion (passing to the corresponding classical-chaos formula in the limit $\hbar \to 0$) [1,17] reproduce respective universal results of sections 2, 3.1 (such as the global chaos criterion of eq. (29)) demonstrating once again their unrestricted universality.

Contrary to the closed system dynamics in the case of Hamiltonian quantum chaos, the case of *quantum measurement* interaction [1,18] involves a small dissipativity of always *quantum* (microscopic) system realising a link to higher, eventually macroscopic levels of measurement device. Therefore instead of performing permanent (frequent) transitions between its well-separated realisations, such slightly dissipative system forms a (transient or stable) multivalued SOC kind of state (see section 3.1), where its wavefunction is reduced (physically squeezed) to a localised configuration around a dissipative "leak" to higher levels, containing many close, practically inseparable realisations (this is the causally explained, physically real "wave reduction"). In the case of transient measurement configuration (as in the double-slit experiment) this unstable self-organised state in then transformed back to the uniform chaos dynamics (well-separated realisations) of free quantum system after the measurement event. In the case of final measurement configuration (as in the Schrödinger-cat kind of experiment), the measured quantum system is "definitely spoiled" by the measurement event and remains in a stable localised (multivalued SOC) configuration after measurement.

This complex dynamics of real quantum measurement provides a good basis for understanding of causal emergence and complex-dynamic origin of *classical behaviour* [1,8,9,13-15,19]. The latter can actually be explained as a permanently localised, multivalued-SOC kind of behaviour of *elementary (closed) bound systems* of quantum elements (particles), such as atoms (and all greater ones). The system should be neither "large enough" nor open to a "decohering" environment, but simply be composed of at least two bound quantum elements. The classical, *permanently localised* kind of behaviour emerges then just due to purely random quantum wandering of virtual solitons of constituent quantum systems. As these quantum beat deviations of bound particles are independent, the probability of respective quantum jumps in one direction (i.e. of the system as a whole) is small and decreases exponentially with the number of "coherent" jumps. The system thus becomes effectively localised due to this limiting link between the elements, even though (but also because) each of them tries to wander quantum-mechanically in an *arbitrary* direction (see e.g. [15], section 1.3.8, for more details). We can naturally explain also, within this description, the effectively quantum behaviour of bound systems with very strong (relativistic) binding interaction (such as hadrons consisting of bound quarks) and quantum behaviour of large enough many-particle systems (remaining puzzling within usual "decoherence" hypotheses) under the influence of suitable external interactions [1,8,9,13-15,19].

**4.4. Emergent universe as a system: Complex-dynamic cosmology**

Complexity unfolding from lower-level interactions to emerging higher-level structures continues in the same, universally specified way up to its highest known levels of living organisms [4,5], human societies and civilisations [22] (including information and communication technologies [23-25]), intelligence and consciousness [21]. This development occurs in a natural irregular alternation of characteristic types of behaviour and dynamical regimes specified above (section 3) and realised already at the lowest complexity levels (section 4), such as global (uniform) chaos and dynamically multivalued SOC, or wave-like ("generalised quantum") and permanently localised ("generalised classical") behaviour, including the effects of "special and general relativity" now extended to any complexity level [1,15]. While we leave the detailed account of these higher-level applications of our unified theory of unreduced interaction complexity (sections 2 and 3) to other papers (see refs. [1-6,21-26]), it would be relevant to summarise here the general *cosmological* results of our approach and respective problem solutions [15,16], as a concrete unifying framework for the entire emerging world dynamics.

Note, first of all, the *intrinsically cosmological* character of our description considering any existing structure as a result of explicit and completely specified interaction development process, where the Universe begins as a global interaction between two primordial protofields (section 4.1). It should be compared to usual theory registration of already existing, basically separated structures which it tries then to unify in a mechanically composed, inevitably deficient cosmological framework. Hence the strangely dominating and ever growing unsolved problems of the latter, such as the notorious dark matter or Big Bang contradictions, despite the apparent modern "triumph" of fundamental science methods and tools.

A mathematically exact and rigorously substantiated summary of this fundamental difference between our complex-dynamic (dynamically multivalued) and unitary (dynamically single-valued) cosmologies is provided by the definitely *positive (and great) value of the universe energy-complexity* in our theory, $E > 0$, vs its zero (or relatively small) value in traditional cosmology. This total energy positivity follows from our universal interaction analysis (see eqs. (34), (36)) as a major manifestation of the fundamental dynamic multivaluedness of interaction results related to the unstoppable and irreversible *time flow* definitely oriented to growing dynamic entropy for *any* real object and process (vs effectively absent or only formally introduced time in usual theory, which gives its well-known and unsolved "problem of time"). This result has therefore a nontrivial origin and universally applicable character reflecting not any detailed quantitative balance of different contributions to the total energy of the universe (as in usual cosmology) but the inevitably dominating part of dynamically random structure creation processes hugely exceeding the artificially reduced dynamic content of conventional one-realisation unitary model with zero value of unreduced dynamic complexity. It is this artificial reduction of real dynamically multivalued world structure to its dynamically single-valued models that is behind all those accumulated difficulties of "missing energy and matter" (as well as missing time) of unitary cosmology that either never exist or find their natural solutions within the unreduced, complex-dynamic description of universe dynamics [15,16].

Another general result of our intrinsically complete many-body interaction description is that we naturally obtain a dynamically adjustable, "fine-tuned" universe (the well-known problem of usual theory) that "tries" automatically to realise all its structure-creation potentialities by fully transforming its dynamic information-complexity into dynamic entropy-complexity [1,15,16] (in particular, due to intrinsic adaptability of probabilistic dynamic fractality of unreduced multivalued dynamics, section 2). It starts specifically from field-particle formation in the interacting protofield system as described above (section 4.1), where the growing number of particle quantum beat processes leads to protofield tension increase until new particles cannot form any more (under average conditions). It means that the total universe mass density and distribution are determined by protofield interaction and naturally attain reasonable well-balanced values, where extreme cases of massless or "too massive" (collapsing) universe represent rather pathological and therefore rare eventualities. The same is true for further structure-formation processes at higher complexity levels, where ever more complicated structure formation always probabilistically wins, with a dynamically determined eventual distribution of results. The entire thus causally emerging universe appears as a single, dynamically unified and time-synchronised structure (section 4.1) of "dynamically multivalued SOC" type (section 3.1) mathematically described by the equally unified dynamically probabilistic fractal (section 2).

One indeed necessary condition for the whole construction to be viable is the existence of the starting protofield system itself with "sufficient" protofield properties (such as great enough elasticity of the e/m protofield). However, as we deal here with the maximum possible simplicity of this initial system configuration, while unique protofield properties are beyond any possible comparison with "similar" entities, starting with this configuration looks not as an excessive but rather as a minimum possible assumption.

The well-known "old" and "new" Big Bang problems do not even appear in our naturally structure-producing description as we simply do not need to evoke any mechanistic "linking" assumptions in order to keep together our intrinsically unified universe structure and dynamics. We obtain from the beginning a perfectly "flat" and physically tangible space with naturally running, equally real but not tangible time that do not need any "expanding" or "squeezing" over-all dynamics as the emerging space structure is "maintained" by innumerable tendencies of multivalued dynamics of all scales more resembling a quasi-permanent (but still fundamentally decaying) "dissipative/turbulent motion" with numerous creation and destruction events than any simple common mechanics underlying usual cosmology framework. The main accepted signature of canonical Big Bang expansion, the famous "red shift" of light quanta frequency at very long propagation distances not only can be explained within the dynamics of coupled protofields but appears as an inevitable dissipation effect, since photons propagated "at the surface" of the strained e/m protofield always preserve their (very weak) interaction with the underlying matrix of the physically real gravitational protofield (realised most probably as a dense quark condensate) and therefore should lose energy to the latter. As to the microwave background radiation, another "definite sign" of the former Big Bang explosion, it appears as inevitable feature of the multivalued protofield interaction dynamics, since the constituting field-particle quantum beat pulsation will always leave enough of "residual trembling" of inter-particle space, where new fully massive particles cannot form any more. The effective "temperature" of this photonic background is determined by thermodynamic considerations, irrespective of other system features (see e.g. [46,47] for details).

The "dark energy" effects including the "accelerated expansion" of the universe observed by red shift variations are explained in our theory in a similar parsimonious way (without evoking additional "invisible" entities) by generally inhomogeneous and nonlinear degradation of photon energy in its long-distance interaction with various inhomogeneous gravitational protofield domains (like those around highly energetic cosmic objects etc.). The observationally different "dark mass" effects (galaxy rotation curves etc.) are explained in a different by generally similar way by multiple realisations of stellar motion components invariably missing in simplified unitary models and artificially replaced by a "visible" influence of additional (but strangely "invisible") matter species or arbitrarily modified Newton's motion law (see [15] for mathematical details). As noted above, all these unitary cosmology deficiencies have the same root of artificially simplified many-body interaction dynamics with its multiple realisations being reduced to a single, "averaged" one.

Let us mention finally various other features and effects that remain unexplained and separated in the unitary theory framework but obtain not only causal but intrinsically unified explanation in our unreduced interaction analysis [1,7-16,19]. These include not only unified explanation of "intrinsic" and "dynamic" particle properties (see above, sections 4.1-4.3) but the "tacit" assumption of their permanence throughout the entire universe, including the unique and synchronised time flow of the universe. As a dynamically single-valued theory cannot properly account for any real change (structure emergence) in principle, it will inevitably encounter particularly difficult, unsolvable problems in consistent explanation of cosmological processes involving essential structure formation dynamics, which are naturally explained by explicitly change-bearing multivalued interaction dynamics giving rise to the physically real time itself.

### 4.5. Experimental confirmation and further development strategy

In this section we shall summarise experimental confirmations and practical consequences of the obtained causally complete picture of complex-dynamic (dynamically multivalued) origin of elementary physical entities, their properties and dynamics in a physically unified process of unreduced interaction between two initially homogeneous protofields (as described above, sections 4.1-4).

(1) <u>Causally explained and unified microworld properties</u>. Many observed and well-established features, properties and laws of fundamental entities (space, time, particles and interactions) remain causally unexplained within usual theory and only formally "postulated", often in the form of a "supernatural", strangely persisting "mystery" or not less physically obscure abstract "principle". Moreover, many of them remain basically separated from one another in origin and properties, sometimes in a highly contradictory way (e.g. canonical "quantum" and "relativistic" properties, interaction forces, etc.), without any unified and physically transparent picture being realistically in view. We now provide a causally complete and intrinsically unified description of observed major structures and features of fundamental physical entities and laws resolving old and new mysteries and contradictions.

(1.1) Among major results we can mention the causally derived, *dynamically emerging* number, origin and properties of tangible space dimensions and irreversibly flowing time, particle structure, species, intrinsic and dynamic properties, including now unified quantum and relativistic behaviour without postulated "mysteries" and "principles", Newton's motion and gravitation laws, number and properties of intrinsically unified fundamental interaction forces causally related to the number and physical origin of space dimensions (sections 4.1, 4.2).

(1.2) We should mention especially the *complex-dynamical origin of inertial mass* intrinsically unified with its gravitational manifestations (sections 4.1, 4.2). A strong practical implication is redundancy of any "model" origin of mass from an additional physical entity (particle species, "hidden dimension" or interaction force) and *uselessness* of experimental search for such entity (such as the abstract and deficient idea of Higgs boson and field), meaning the necessity of qualitative strategy change in today's experimental fundamental physics (see also below).

(1.3) Similar practical conclusion follows from another cornerstone of modern experimental searches, the (conventional) Planckian units and related hierarchy problem. Our *causally renormalized Planckian units* (section 4.2) show, without evoking any inconsistently "hidden" and abstract entities, that the "extreme" values of mass, length and time interval of this world are already (approximately) attained in the observed particle species or at least any essentially more extreme values would be definitely redundant for the observed world construction. Therefore there is no sense to experimentally search for those "harder" species implied by conventional Planckian units as it is either useless to "count" on those grossly exaggerated values in various theoretical models (as actually very widely done in various directions of unitary theory, from string theory and quantum gravity to cosmology, becoming thus additionally compromised).

(1.4) We can also mention in the same category of now definitely useless and practically harmful abstract constructions the well-known idea of "supersymmetry" between bosons and fermions occupying a prominent place in conducted experimental search of industrial scale. In our causal microworld picture we reveal the real physical origin and dynamic structure of all particles (contrary to their purely abstract presentation in usual theory), including "interaction exchange" particles (such as photons) and other bosonic species, which shows their real physical nature and difference from fermionic species [1] thus leaving no place or necessity to any "supersymmetric partners". Various related "false infinities" in abstract calculations of usual theory do not even appear in our physically based description (such as massive "virtual particles", violating the universal complexity conservation law).

(1.5) One may also emphasise the physically and mathematically complete origin and dynamic meaning of the main physical constants, such as $\hbar$ (Planck's constant), $c$ (the speed of e/m waves propagation), $\gamma$ (gravitational constant), $e$ (elementary charge) and $\alpha$ (fine-structure constant), revealed in our theory (sections 4.1, 4.2), contrary to their purely abstract role of postulated "coefficients" in unitary theory. Although it doesn't directly imply experimental novelties, these important quantities, provided now with their physically complete meaning and relations, demonstrate convincingly the causal completeness of the whole emerging world construction suggesting objectively efficient strategy of its further exploration.

(1.6) A new separate correlation of "unified" physical nature that could not appear in usual theory is the fact of (approximately) *equal maximum masses of elementary particle and atomic nucleus* (of the order of 100 GeV) [9,11,15,16]. Whereas it's nothing but a coincidence within a usual, empirically based description, it cannot be so in our dynamically unified picture where a nucleus with its highly coherent internal dynamics can be considered as a pathologically big and internally complicated particle limited, as such, by the same value of maximum local "strength" of protofield interaction preserving their integrity. While even simplest ele-mentary particles possess, within our interaction analysis, complex (multi-valued) internal dynamics, this correlation reveals the previously unexpected ultimate, "truly fundamental" and therefore quite simple limit to heavy nuclei stability, with due implication for respective experimental searches in nuclear physics. This intrinsic unity of externally different cases of a single elementary particle and their dense agglomeration in atomic nucleus is further supported by universal applicability to both cases of the same quantum laws and Planck's constant, obtaining now its causal explanation (section 4.2). Moreover, it becomes evident that many *high-energy scattering features* so intensely explored within the ongoing LHC adventure may actually witness various dissociation channels of that unified maximum-squeeze state of the highest protofield attraction magnitude (just around $E = M_\text{P} c^2 \simeq 10^2$-$10^3$ GeV, see eq. (65)), rather than of any additional particle species (Higgs etc.).

(2) Special experimental confirmation of underlying interaction dynamics. Whereas usual theory traditionally relies on empirically based postulation of all major entities in the form of related abstract symbols and rules, our analysis of unreduced many-body interaction emphasises a direct dynamic origin and explicit emergence of all observed entities, properties and laws. Apart from correlations in the obtained interaction results mentioned above, one may get therefore some more special and direct experimental signs of fundamental interaction processes involved.

(2.1) Whereas the detailed quantum beat dynamics and its separated phases cannot be directly traced experimentally (because it forms the very lowest complexity sublevel of this world), the existence of quantum beat pulsation as a whole (first assumed by Louis de Broglie and used in the original derivation of his famous expression for the particle wavelength [37]) can be registered experimentally and was actually traced by resonance with periodic collisions of relativistic channelled electrons with crystal lattice atoms [48,49]. While our quantum channeling description within the same unreduced interaction analysis [28] confirms the observed effect and could be used for its detailed analysis, these fundamentally important but occasional experiments haven't received any development (to be compared with huge but vain experimental efforts to confirm usual theory assumptions, as discussed above, items 1.2, 1.3). Various other resonance effects revealing the reality of complex quantum-beat dynamics can be expected (see [13], section 3, item (7)) and are waiting for their experimental observation.

(2.2) Recent discovery of the properties of dense quark-gluon *liquid* kind of behaviour in high-energy collision experiments [50] (instead of quark-gluon plasma expected from usual theory) is a qualitative but strong argument in favour of our picture, with its "gravitational protofield" being represented by a dense quark condensate. Additional related facts confirming the emerging interaction configuration is the absence of strong interaction for leptons (a mere empirical fact in usual theory) and the famous quark "confinement" remaining physically obscure in the standard theory framework. We see now that quarks appearing in various high-energy interactions do not emerge simply "from vacuum" due to energy conservation law (where the unreduced complexity conservation law may still be violated) but come from the *omnipresent* gravitational protofield, in the form of inevitably quantised excitations, or "droplets" (starting from two or three quarks in size), of the dense liquid of its ground-level condensate. No particle motion and interaction in this world can happen without involvement of this omnipresent but directly unobservable, effectively liquid condensate of gravitational protofield, which provides a natural explanation for a large scope of observations.

(2.3) As a summary of particle-physics experimental confirmations and perspectives of items (1)-(2), one must emphasise the necessity of a *crucial transition in high-energy research strategy*, from today's blind search in the direction of *quantitative* energy and intensity growth vaguely guided (but actually mislead) by deficient abstract models and giving no constructive results any more to consistently substantiated exploration of *new qualities* of *complex* and *unified* microworld dynamics (basically within the attained quantitative ranges), with further promising applications of both fundamental and practical importance.

(3) <u>Quantum chaos, quantum measurement and classicality emergence</u>. Although we have provided here only a brief account of the results of our theory application at these higher but still basically "quantum" complexity sublevels of interacting elementary particles (section 4.3), it's worthwhile to mention the related practically important and transparent implications for fundamental and applied research in various fields. In fact, this group of applications on the border between quantum and classical world can be considered as the final confirmative "closure" of this whole group of fundamental microworld applications of unreduced many-body interaction complexity.

(3.1) The *quantum chaos* case of Hamiltonian (closed-system) interaction in practically any real quantum system demonstrates the power of our approach to solve the respective persisting fundamental problem (with numerous applications) due to the proposed explicitly extended, dynamically multivalued analysis (see section 4.3 and papers [1,15,17,19] for details and references). We qualitatively evolve here from unpleasantly absent (or artificially imitated) dynamic quantum randomness in usual theory to our *genuine quantum chaoticity* with straightforward transition to classical chaos results, in agreement with the canonical *correspondence principle* (now extended thus to chaotic systems). Application to particle channeling [28] is of particular interest due to its relation to experimental confirmation of quantum beat dynamics (see item (2.1) above).

(3.2) The *quantum measurement* case is but a slightly dissipative version of Hamiltonian quantum chaos of the previous item but with a considerable change in complex interaction results (section 4.3). We obtain here totally causal solution to respective traditional problems, now without any "mysteries" but involving instead a physically real transient localisation of the system and its usually extended wavefunction, in full agreement with observations and postulated empirical rules [1,18].

(3.3) *Classicality* emerges in our theory as the next higher level of our universally defined complexity (section 2), the one of permanently localised behaviour of *elementary bound systems*, such as atoms (and more complex systems), section 4.3. Although such simplest classical systems need not be absolutely closed, the classical, permanently localised behaviour will appear even in an absolutely closed (and microscopic) bound system without any "decoherence" effects but due to the essentially probabilistic (dynamically multivalued) internal structure of such elementary self-organised interaction process [1,8,9,13-15,19]. It appears simply as a result of strong enough, binding interaction as opposed to non-binding (smaller or repulsive) interaction magnitudes in quantum measurement and quantum chaos situations (items (3.1), (3.2)) with only transient binding effects. The respective "closing" mystery of usual quantum mechanics at the border with classical world is thus consistently resolved, including all related applications, such as quantum behaviour revival in various macroscopic "condensates" and for unusually heavy many-body molecular species in suitable additional interactions (which is difficult to understand within the unitary decoherence concept).

(3.4) All these situations of complex quantum interaction dynamics near the border between quantum and emerging classical behaviour, items (3.1)-(3.3), appear in the vast scope of applications of various *atomic-size functional structures* and *quantum machines* practically dealt with in

*nano(bio)technology* [19,20]. Instead of usual empirically based approach, we can propose now a causally complete description of the unreduced complex dynamics of real interactions involved. Moreover, we rigorously show that such explicitly complex (dynamically multivalued) effects as global (uniform) chaos cannot be avoided just at this small, atomic scale of essential interactions involved, implying qualitative deficiency of any unitary, basically regular model.

(3.5) A key feature of unreduced many-body interaction processes with really many interacting entities is their *exponentially huge efficiency* (section 2) inevitably neglected within usual dynamically single-valued models and opening strong application possibilities, in particular at these lowest, "quantum" complexity levels [5,6,19-25]. By contrast, its unitary imitation by expected high and largely mystified efficiency of "quantum computers" cannot be realised just because of inevitable chaoticity of real interactions [19], with obvious implications for technology development.

(3.6) The same application group includes elementary *biological systems* (eventually in their combination with artificial nano-structures). The above feature of huge exponential efficiency of unreduced many-body interaction takes here the form of "magic" properties of *life* remaining basically unexplained in unitary science and now causally understood as high enough level of dynamic complexity of unreduced, dynamically multivalued interaction [4,5] (according to the universal complexity definition of section 2). Further biological applications include unreduced interaction analysis of genome dynamics leading to *causally complete genomics* and very important limitations of usual, empirical genetics [5] that merit a separate detailed presentation.

(3.7) Finally one may mention within the same group various stagnating "difficult" cases of many-body interaction and *solid-state theory*, such as high-temperature superconductivity and various *strong-interaction* cases in general. These remain potential applications to be yet realised but which will certainly need the unreduced, dynamically multivalued interaction analysis providing already the observed qualitative properties and a well-specified origin of strangely persisting difficulties of usual, dynamically single-valued models.

(4) <u>Complex-dynamic cosmology</u>. Another "embracing summary" of fundamental applications of unreduced interaction description appears on the opposite extreme scale of entire universe in the form of causally complete and naturally creative cosmology essentially extending its strongly deficient unitary models (section 4.4) [15,16].

(4.1) Both *old and new problems* of traditional Big Bang cosmology are naturally solved (and often do not even appear) in our intrinsically creative, structure-forming interaction analysis of the entire universe system. We provide the unified origin of usual theory difficulties in the form of "missing realisations" (dynamically single-valued simplification) of real interaction processes and its universal extension to the unreduced multivalued dynamics with strictly positive (and large) total complexity-energy [15,16]. While the growing difficulties of traditional Big Bang solutions are well known and increasingly discussed, we provide their common origin and especially missing unified solution of all these and other fundamental problems going thus far beyond any unitary model.

(4.2) Our causally complete cosmology picture provides a clear demonstration of redundant, unnecessary character of various additional entities of usual dynamically single-valued models introduced artificially in order to compensate (as we can see now) the missing natural richness of unreduced, dynamically multivalued interaction dynamics on the scale of universe. These entities include various multiplying versions of purely abstract "hidden dimensions", "dark matter" and "dark energy" species (particles and fields) of "invisible" nature but quite visible manifestations (exactly where necessary), or else arbitrary formal modifications of major laws of Newton dynamics and gravitation. By providing the universal and consistent argument against their existence, we initiate successfully the badly needed work towards essential increase of efficiency of (very expensive) experimental research in cosmology as well as the reliable basis for its future creative strategy.

In summary of section 4.5, the system of main experimental confirmations and practical development perspectives grouped in above 18 items (1)-(4) provides convincing support for our theoretical microworld description in terms of complex (multivalued) many-body interaction dynamics. Practically important and efficient applications of the same causally complete interaction analysis continue to all higher complexity levels with remarkable permanent recurrence of "quantum" and "relativistic" manifestations of dynamic complexity at those higher levels of unreduced world dynamics [1,6,15] demonstrating once again their genuine, complex-dynamic origin. While each of these applications would need a separate description, all of them are characterised by a *single unified structure* of fundamental *dynamically probabilistic fractal* of all world interactions (section 2) and the *single unifying principle* of the *universal symmetry (conservation and transformation) of complexity* underlying all particular (correct) laws (section 3.2), which is a well-specified *new mathematics of* *complexity and emergence* [1,6,15,22,23].

In conclusion of section 4 (fundamental applications of complex-dynamic interaction analysis) one may add that the above results certainly suppose further development in various directions of particle physics and cosmology, including introduction of really indispensable new entities. We have only shown here that all traditional "mysteries", persisting old and emerging new problems of fundamental physics can be causally, realistically resolved without introduction of new entities but only due to the truly complete solution of the unreduced many-body interaction problem giving rise to a huge variety of system realisations and related dynamic regimes appearing in observed effects but artificially missing from usual, dynamically single-valued approximation. Therefore this is precisely the new mathematics of complexity needed to upgrade the artificially limited unitary science basis up to its causally complete, totally realistic version liberated from supernatural mysteries and unsolvable problems.

This is the essential difference of the proposed *unified* solution to fundamental (including quantum) problems [1,7-19] from various unitary "models" of separate, isolated features appearing increasingly since then (e.g. [51-66]) that also often refer to *assumed* "hidden dynamics" or ill-defined "complexity" but do not specify either their real, causal origin or the key novelty behind the expected new understanding (such attempts are often reduced to variations of known, always incomplete quantum "interpretations", such as Nelson's stochastic dynamics [67] or complementary "undular" models, whereas our description can be considered as the dynamically multivalued extension of the full "double solution" of Louis de Broglie [36,37,68-70]). As a result, those invariably unitary (dynamically single-valued) models of dynamically multivalued reality are fatally deficient in their emphasis of certain (e.g. stochastic) aspects of complex dynamics and inevitable absence of other, equally important aspects (like undular and regular motion components). Unitary models can reproduce only structures (and equations) that were actually postulated from the beginning. We have proposed the needed causal origin of *explicit structure emergence* in the form of unreduced interaction process providing the qualitative novelty of dynamically multivalued (redundant) interaction result [1-26,28] and demonstrated the unrestricted universality of thus obtained complexity definition in a vast variety of applications starting from fundamental physics reviewed here.

# REFERENCES

1. A.P. Kirilyuk, *Universal Concept of Complexity by the Dynamic Redundance Paradigm: Causal Randomness, Complete Wave Mechanics, and the Ultimate Unification of Knowledge* (Kyiv: Naukova Dumka: 1997). For a non-technical review see also ArXiv:physics/9806002.
2. A.P. Kirilyuk, *Solid State Phenomena*, **97-98**: 21 (2004). ArXiv:physics/0405063.
3. A.P. Kirilyuk, *Proceedings of Institute of Mathematics of NAS of Ukraine*, **50**, Part 2: 821 (2004). ArXiv:physics/0404006.
4. A.P. Kirilyuk, In: G.A. Losa, D. Merlini, T.F. Nonnenmacher, and E.R. Weibel (Eds.), *Fractals in Biology and Medicine, Vol. III* (Basel: Birkhäuser: 2002), p. 271. ArXiv:physics/0305119.
5. A.P. Kirilyuk, In: G.A. Losa, D. Merlini, T.F. Nonnenmacher, and E.R. Weibel (Eds.), *Fractals in Biology and Medicine, Vol. IV* (Basel: Birkhäuser: 2005), p. 233. ArXiv:physics/0502133.
6. A.P. Kirilyuk, "Universal Science of Complexity: Consistent Understanding of Ecological, Living and Intelligent System Dynamics", *ArXiv:0706.3219*.
7. A.P. Kirilyuk, "Double Solution with Chaos: Dynamic Redundance and Causal Wave-Particle Duality", *ArXiv:quant-ph/9902015*.
8. A.P. Kirilyuk, "Double Solution with Chaos: Completion of de Broglie's Nonlinear Wave Mechanics and Its Intrinsic Unification with the Causally Extended Relativity", *ArXiv:quant-ph/9902016*.
9. A.P. Kirilyuk, "Universal gravitation as a complex-dynamical process, renormalised Planckian units, and the spectrum of elementary particles", *ArXiv:gr-qc/9906077*.
10. A.P. Kirilyuk, "75 Years of Matter Wave: Louis de Broglie and Renaissance of the Causally Complete Knowledge", *ArXiv:quant-ph/9911107*.
11. A.P. Kirilyuk, "100 Years of Quanta: Complex-Dynamical Origin of Planck's Constant and Causally Complete Extension of Quantum Mechanics", *ArXiv:quant-ph/0012069*.
12. A.P. Kirilyuk, "75 Years of the Wavefunction: Complex-Dynamical Extension of the Original Wave Realism and the Universal Schrödinger Equation", *ArXiv:quant-ph/0101129*.
13. A.P. Kirilyuk, "Quantum Field Mechanics: Complex-Dynamical Completion of Fundamental Physics and Its Experimental Implications", *ArXiv:physics/0401164*.
14. A.P. Kirilyuk, "Electron as a Complex-Dynamical Interaction Process", *ArXiv:physics/0410269*.
15. A.P. Kirilyuk, "Consistent Cosmology, Dynamic Relativity and Causal Quantum Mechanics as Unified Manifestations of the Symmetry of Complexity", *ArXiv:physics/0601140*.
16. A.P. Kirilyuk, "Complex-Dynamical Approach to Cosmological Problem Solution", *ArXiv:physics/0510240*.
17. A.P. Kirilyuk, *Annales de la Fondation Louis de Broglie*, **21**, Iss. 4: 455 (1996). ArXiv:quant-ph/9511034, quant-ph/9511035, quant-ph/9511036.
18. A.P. Kirilyuk, "Causal Wave Mechanics and the Advent of Complexity. IV. Dynamical origin of quantum indeterminacy and wave reduction", *ArXiv:quant-ph/9511037*.
19. A.P. Kirilyuk, "Dynamically Multivalued, Not Unitary or Stochastic, Operation of Real Quantum, Classical and Hybrid Micro-Machines", *ArXiv:physics/0211071*.

20. A.P. Kirilyuk, *Nanosystems, Nanomaterials, Nanotechnologies*, **2**, Iss. 3: 1085 (2004). ArXiv:physics/0412097.
21. A.P. Kirilyuk, "Emerging Consciousness as a Result of Complex-Dynamical Interaction Process", *ArXiv:physics/0409140*.
22. A.P. Kirilyuk, In: V. Burdyuzha (Ed.), *The Future of Life and the Future of Our Civilisation, Vol. IV* (Dordrecht: Springer: 2006), p. 411. ArXiv:physics/0509234.
23. A.P. Kirilyuk, In: D. Gaïti (Ed.), *Network Control and Engineering for QoS, Security, and Mobility, IV, IFIP, Vol. 229* (Boston: Springer: 2007), p. 1. ArXiv: physics/0603132.
24. A. Kirilyuk and M. Salaün, In: D. Gaïti (Ed.), *Network Control and Engineering for QoS, Security, and Mobility, V, IFIP, Vol. 213* (Boston: Springer: 2006), p. 223.
25. A. Kirilyuk and M. Ulieru, In: M. Ulieru, P. Palensky and R. Doursat (Eds.), *IT Revolutions* (Berlin Heidelberg: Springer: 2009), p. 1. ArXiv:0910.5495.
26. A.P. Kirilyuk, In: M. López Corredoira and C. Castro Perelman (Eds.), *Against the Tide: A Critical Review by Scientists of How Physics & Astronomy Get Done* (Boca Raton: Universal Publishers: 2008), p. 179. ArXiv:0705.4562.
27. P.H. Dederichs, In: H. Ehrenreich, F. Seitz and D. Turnbull (Eds.), *Solid state physics: Advances in research and applications, vol 27* (New York: Academic Press: 1972), p. 136.
28. A.P. Kirilyuk, *Nucl. Instr. and Meth.*, **B69**, Iss. 2-3: 200 (1992).
29. R. Penrose, *Shadows of the Mind* (New York: Oxford University Press: 1994).
30. L.D. Landau and E.M. Lifshitz, *Mechanics* (Moscow: Nauka: 1988). Forth Russian edition.
31. H. Haken, *Advanced Synergetics* (Berlin: Springer-Verlag: 1983).
32. B.V. Chirikov, *Physics Reports*, **52**, Iss. 5: 263 (1979).
33. A.J. Lichtenbegr and M.A. Lieberman, *Regular and Stochastic Motion* (New York: Springer-Verlag: 1983).
34. G.M. Zaslavsky, *Chaos in Dynamical Systems* (London: Harwood Academic Publishers: 1985). Russian edition: Moscow: Nauka: 1984.
35. H.G. Schuster, *Deterministic Chaos* (Weinheim: Physik-Verlag: 1984).
36. L. de Broglie, *La thermodynamique de la particule isolée (thermodynamique cachée des particules)* (Paris: Gauthier-Villars: 1964).
37. L. de Broglie, *Annales de Physique (10e Série)*, **III**: 22 (1925). Reprinted edition: L. de Broglie, *Recherches sur la théorie des quanta* (Paris: Fondation Louis de Broglie: 1992).
38. L.D. Landau and E.M. Lifshitz, *Field theory* (Moscow: Nauka: 1989). Sixth Russian edition.
39. J.-M. Raimond and V. Rivasseau (Eds.), *Quantum Decoherence* (Basel: Birkhäuser: 2006).
40. A. Bassi and G.C. Ghirardi, *Physics Reports*, **379**, Iss. 5-6: 257 (2003). ArXiv: quant-ph/0302164.
41. R. Omnès, "Decoherence and wave function collapse", *ArXiv:1105.0831*.
42. W.H. Zurek, *Rev. Mod. Phys.* **75**, Iss. 3: 715 (2003). ArXiv:quant-ph/0105127.
43. W.H. Zurek, *Nature Physics*, **5**: 181 (2009). ArXiv:0903.5082.
44. J.P. Paz and A.J. Roncaglia, *Phys. Rev. A*, **80**, Iss. 4: 042111 (2009). ArXiv: 0909.0474.
45. L. Diósi, *J. Phys: Conf. Ser.*, **174**: 012002 (2009). ArXiv:0902.1464.
46. R.E. Criss and A.M. Hofmeister, *Geochimica et Cosmochimica Acta*, **65**, Iss. 21: 4077 (2001).

47. J.G. Hartnett, "Is the Universe really expanding?", *ArXiv:1107.2485*.
48. M. Gouanère, M. Spighel, N. Cue, M.J. Gaillard, R. Genre, R. Kirsch, J.C. Poizat, J. Remillieux, P. Catillon and L. Roussel, *Annales de la Fondation Louis de Broglie*, **30**, Iss. 1: 109 (2005).
49. G. Lochak, *Annales de la Fondation Louis de Broglie*, **30**, Iss. 1: 115 (2005).
50. J. Adams et al. (STAR Collaboration), *Nucl. Phys. A*, **757**: 102 (2005). ArXiv:nucl-ex/0501009.
51. A. Rueda and B. Haisch, *Phys. Lett. A*, **240**, Iss. 3: 115 (1998). ArXiv: physics/9802031.
52. A. Rueda and B. Haisch, *Phys. Lett. A*, **268**, Iss. 4-6: 224 (2000). ArXiv:gr-qc/9906084.
53. A. Khrennikov, *J. Phys. A: Math. Gen.*, **38**, Iss. 41: 9051 (2005).
54. A. Khrennikov, *Found. Phys. Lett.*, **18**, Iss. 7: 637 (2005).
55. E. Verlinde, *JHEP*, **2011**, Iss. 4: 29 (2011). ArXiv:1001.0785.
56. A. Caticha, *J. Phys. A: Math. Theor.*, **44**, Iss. 22: 225303 (2011). ArXiv:1005.2357.
57. G. Grössing, S. Fussy, J. Mesa Pascasio and H. Schwabl, *Physica A*, **389**, Iss. 21: 4473 (2010). ArXiv:1004.4596.
58. V. Allori, S. Goldstein, R. Tumulka and N. Zanghì, *Brit. J. Philos. Sci.*, **59**: Iss. 3: 353 (2008). ArXiv:quant-ph/0603027.
59. R. Sverdlov, "Can Bohmian particle be a source of "continuous collapse" in GRW-type theories", *ArXiv:1103.2889*.
60. D.J. Bedingham, *J. Phys. A: Math. Theor.*, **44**, Iss. 27: 275303 (2011). ArXiv: 1104.1938.
61. S. Colin and H. M. Wiseman, *J. Phys. A: Math. Theor.*, **44**, Iss.34: 345304 (2011). ArXiv:1107.4909.
62. S. Gao, *AIP Conf. Proc.*, **1327**, Iss. 1: 334 (2011). ArXiv:1108.1187.
63. W.A. Hofer, *Found. Phys.*, **41**, Iss. 4: 754 (2011). ArXiv:1002.3468.
64. L. Randall and R. Sundrum, *Phys. Rev. Lett.*, **83**, Iss. 17: 3370 (1999). ArXiv:hep-ph/9905221.
65. L. Randall and R. Sundrum, *Phys. Rev. Lett.*, **83**, Iss. 23: 4690 (1999). ArXiv:hep-th/9906064.
66. J. Khoury, B. A. Ovrut, P. J. Steinhardt and N. Turok, *Phys. Rev. D*, **64**, iss.12: 123522 (2001). ArXiv:hep-th/0103239.
67. E. Nelson, *Phys. Rev.*, **150**, Iss. 4: 1079 (1966).
68. L. de Broglie, *Une tentative d'interprétation causale et non-linéaire de la mécanique ondulatoire: la théorie de la double solution* (Paris: Gauthier-Villars: 1956). English translation: *Nonlinear Wave Mechanics* (Amsterdam: Elsevier: 1960).
69. L. de Broglie, *Le Journal de Physique et le Radium*, **20**, Iss. 12: 963 (1959).
    L. de Broglie, *Annales de la Fondation Louis de Broglie*, **22**, Iss. 4: 336 (1997).
70. A.P. Kirilyuk, "75 Years of Matter Wave: Louis de Broglie and Renaissance of the Causally Complete Knowledge", *ArXiv:quant-ph/9911107*.